\documentclass[sigconf]{acmart}

\usepackage{booktabs} % To thicken table lines

\usepackage[english]{babel}
\usepackage{moresize}
\usepackage{amsmath}
\usepackage{algorithmic} 
\usepackage{balance}
\usepackage{comment}
\usepackage{paralist}
\usepackage{bm}
\usepackage{pgfplots}
\usetikzlibrary{pgfplots.dateplot}

\usepackage{flushend}
\usepackage[english]{babel}
\usepackage[latin1]{inputenc}
\usepackage{mathrsfs}
\usepackage{graphicx}
\usepackage{amsfonts}
\usepackage{url}
\usepackage{longtable}
\usepackage{rotating}
\usepackage{multirow}
\usepackage{mathrsfs}
\usepackage{subfigure}
\usepackage{enumitem}
\usepackage[linesnumbered,algoruled,boxed,lined]{algorithm2e}
\usepackage{adjustbox}
\usepackage{hyperref}

\usepackage{pgfplots}
\usetikzlibrary{pgfplots.dateplot}

\usepackage{filecontents}
\definecolor{tblue}{RGB}{31,119,180}
\definecolor{torange}{RGB}{255,127,14}
\definecolor{tgreen}{RGB}{44,160,44}
\definecolor{tred}{RGB}{214,39,40}
\definecolor{tpurple}{RGB}{148,103,189}

\newcommand{\hide}[1]{} %hide

\newcommand{\etal}{\textit{et al}.}

\newcommand{\ie}{\textit{i}.\textit{e}.}
\newcommand{\eg}{\textit{e}.\textit{g}.} 
\newcommand{\wrt}{\textit{w}.\textit{r}.\textit{t}} 
\newtheorem{Dfn}{Definition}

\setcopyright{none}
\copyrightyear{2020}
\acmYear{2020}
\setcopyright{acmcopyright}\acmConference[SIGIR '20]{Proceedings of the 43rd International ACM SIGIR Conference on Research and Development in Information Retrieval}{July 25--30, 2020}{Virtual Event, China}
\acmBooktitle{Proceedings of the 43rd International ACM SIGIR Conference on Research and Development in Information Retrieval (SIGIR '20), July 25--30, 2020, Virtual Event, China}
\acmPrice{15.00}
\acmDOI{10.1145/3397271.3401445}
\acmISBN{978-1-4503-8016-4/20/07}

\def\model{MATN}

\begin{document}

\fancyhead{}

\title{Multiplex Behavioral Relation Learning for Recommendation via Memory Augmented Transformer Network}

\author{Lianghao Xia}
\authornote{These authors contributed equally to this work.}
\affiliation{South China University of Technology}
\email{cslianghao.xia@mail.scut.edu.cn}

\author{Chao Huang}
\affiliation{\institution{JD Finance America Corporation}}
\authornotemark[1]
\email{chaohuang75@gmail.com}

\author{Yong Xu}
\authornote{Corresponding author}
\affiliation{\institution{South China University of Technology\\Peng Cheng Laboratory}}
\email{yxu@scut.edu.cn}

\author{Peng Dai}
\affiliation{\institution{JD Finance America Corporation}}
\email{peng.dai@jd.com}

\author{Bo Zhang}
\affiliation{\institution{Boshi Qiangzhi Science and Technology Co., Ltd}}
\email{13922820911@139.com}

\author{Liefeng Bo}
\affiliation{\institution{JD Finance America Corporation}}
\email{liefeng.bo@jd.com}

\begin{abstract}
Capturing users' precise preferences is of great importance in various recommender systems (\eg, e-commerce platforms and online advertising sites), which is the basis of how to present personalized interesting product lists to individual users. In spite of significant progress has been made to consider relations between users and items, most of existing recommendation techniques solely focus on singular type of user-item interactions. However, user-item interactive behavior is often exhibited with multi-type (\eg, page view, add-to-favorite and purchase) and inter-dependent in nature. The overlook of multiplex behavior relations can hardly recognize the multi-modal contextual signals across different types of interactions, which limit the feasibility of current recommendation methods. To tackle the above challenge, this work proposes a \underline{\textbf{M}}emory-\underline{\textbf{A}}ugmented \underline{\textbf{T}}ransformer \underline{\textbf{N}}etworks (\model), to enable the recommendation with multiplex behavioral relational information, and joint modeling of type-specific behavioral context and type-wise behavior inter-dependencies, in a fully automatic manner. In our \model\ framework, we first develop a transformer-based multi-behavior relation encoder, to make the learned interaction representations be reflective of the cross-type behavior relations. Furthermore, a memory attention network is proposed to supercharge \model\ capturing the contextual signals of different types of behavior into the category-specific latent embedding space. Finally, a cross-behavior aggregation
component is introduced to promote the comprehensive collaboration across type-aware interaction behavior representations, and discriminate their inherent contributions in assisting recommendations. Extensive experiments on two benchmark datasets and a real-world e-commence user behavior data demonstrate significant improvements obtained by \model\ over baselines. Codes are available at: https://github.com/akaxlh/MATN.
\end{abstract}




\maketitle

\section{Introduction}
\label{sec:intro}

Recommender system, which facilitates the information-seeking process of users and meet their personalized interests, have played a critical role in various online services, such as e-commerce systems~\cite{wang2020time,huang2019online}, online review platforms~\cite{altenburger2019yelp,2019neuraltensor} and advertising~\cite{wang2019learning}. At its core is to learn low-dimensional representations of user-item interaction while capturing the user preference and the underlying intrinsic characteristics~\cite{liu2019compositional}. Early methods towards this goal, have made significant efforts on transforming user-item interactions through vectorized representations based on the conventional Collaborative Filtering (CF) techniques (\eg, matrix factorization scheme~\cite{koren2009matrix,mnih2008probabilistic} and its variations~\cite{rendle2012bpr,he2016fast}).

Inspired by the advancement of deep learning techniques, various neural network-based collaborative filtering frameworks have been developed to model the relationships between users and items. These methods aim to map sparse input interaction features into low-dimensional user/item embedding vectors and then project them into fixed-length representations in a group-wise manner~\cite{zhou2018deep,li2019multi}. For example, neural collaborative filtering models replace the inner product function in the matrix factorization consider non-linearities with multilayer perceptron~\cite{he2017neuralncf} and metric learning scheme~\cite{tay2018latent}. In addition, auto-encoder architecture has served as an effective solution to learn a mapping function between the explicit interaction and latent representation through the reconstruction-based encoder-decoder framework. To capture the rich graph-based neighborhood contextual signals, various graph neural encoders have been proposed to aggregate information over the user-item interaction graph, with the graph convolutional network~\cite{zhang2019star} or message-passing mechanism~\cite{wang2019neural}. 

Despite the prevalence of the above recommendation solutions, they has thus far focused on user preference representation learning with the consideration of singular type of user-item interactive behavior. However, in many practical recommendation scenarios, user-item interactions are multiplex and exhibited with relationship diversity in nature. Let's consider the e-commerce system as an example, there exist multiple types of behavior (\eg, page view, add-to-favourite, add-to-cart and purchase) between users and items~\cite{guo2019buying}, which are mutually inter-dependent. For instance, add-to-cart behavior is more likely to co-occur with purchase than the add-to-favorite behavior. The page view and add-to-favourite behavior can also provide useful signals for making purchase decisions. In such cases, the ignorance of such multi-modal relations across different types of user-item interaction behavior, makes existing recommendation methods insufficient to distill effective collaborative signals from the collective users behavior.

The recommendation framework with multiplex interactive behavior pose two key challenges: \emph{First}, the dependencies across different types of user-item interactions can be arbitrary since any pair of type-specific behavior could potentially be correlated due to various factors~\cite{wen2019leveraging}. For example, users often have correlated online behaviors and exhibit different dependencies in choosing items of different categories due to his/her specialty. Such inter-dependencies between different types of interaction behavior may vary by users and items. While a handful of studies attempt to learn user preference from multi-behavior~\cite{gao2019neural,guo2019buying}, they merely consider the singular dimensional cascading correlations between multi-type interactions, and cannot comprehensively capture the arbitrary dependencies between different types of interaction over different items. Hence, to build effective recommendation model with the complex behavior dependencies remains a significant challenge.

\emph{Second}, when modeling the relationships across different types of behavior, it is also important to capture the context and semantics of individual type of user-item interactions, \eg, users' page view are more frequent than their purchases, and add-to-favorite behavior is more likely to happen over users' interested items but may postpone their buying decision. In addition, type-specific behavioral patterns interweave with each other in complex way (\eg, support or mutually exclusive relations) and are difficult to be captured. During the behavioral pattern integration, as the importance of various types of behavior can be different, their relevance in assisting the forecasting task on the target behavior need to be carefully decided.

Motivated by the aforementioned challenges, this work proposes a general and flexible multi-behavior relation learning framework--\underline{\textbf{M}}emory-\underline{\textbf{A}}ugmented \underline{\textbf{T}}ransformer \underline{\textbf{N}}etworks (\model). Specifically, this work first proposes a multi-behavior transformer network to learn type-specific behavioral representations with the incorporation of inter-dependencies across different types of user-item interactions. By integrating the transformer network with a memory-augmented attention mechanism, we endow the \model\ framework with the capability of incorporating type-specific behavior contextual signals. to collectively model the implicit relevance across multi-type behavioral patterns and perform comprehensive learning for making recommendations, we design a behavior type-wise gating mechanism which promotes the collaboration of different types of interactions. In the pattern aggregation layer, \model\ could learn cross-type representations in the latent feature spaces by automatically adjusting the contribution of each behavior view point in the behavior predictive model.

% To support the reproducibility of the results in this study, we have released our source code and data\footnote{https://github.com/anonymous0402/MATN}.

The contributions of this paper are highlighted as follows:

\begin{itemize}[leftmargin=*]
\item We propose \model, a new recommendation framework with multiplex behavioral relation learning. \model\ explicitly encodes multi-behavior relational structures by preserving both the cross-type behavior collaborative signals and type-specific behavior contextual information. \\\vspace{-0.1in}

\item We first develop a multi-behavior dependency encoder with a transformer architecture, to inject collaborative signals across different types of user-item interactions into the embedding process. Furthermore, we augment the multi-behavior transformer network with a memory attention mechanism, which is capable of uncovering type-specific behavior semantics during the customized representation recalibration phase. \\\vspace{-0.1in}

\item Finally, a type-wise pattern aggregation layer with gating mechanism is developed to promote the collaboration of different behavior views for robust representations on user preferences. \\\vspace{-0.1in}

\item Our extensive experiments on two benchmark datasets and a user behavior data from a major e-commence platform, demonstrate that \model\ outperforms 12 baselines from various research lines in yielding better recommendation performance. We further perform case studies with qualitative examples to better understand the interpretation ability of \model\ framework, and study the model efficiency under different recommendation scenarios.
\end{itemize}

\section{Preliminary}
\label{sec:model}

In the recommendation scenario, we first define the behavior (\eg, purchase) which we aim to predict as \emph{target behavior}, other relevant user-item interactive behavior (\eg, click, add-to-cart and add-to-favorite) is termed as \emph{source behavior}. In this work, we aim to explore the latent relational structures between different types of user behavior (\eg, purchases and click) for making predictions on the target behavior of users in recommender systems.

\begin{Dfn}
\noindent \textbf{Multi-Behavior Tensor $\mathcal{\boldmath{X}}$}. We define a three-dimensional multi-behavior tensor $\mathcal{\boldmath{X}} \in \mathbb{R}^{I\times J\times L}$ to represent the $L$ (indexed by $l$) types of behavior from $I$ (indexed by $i$) users over $J$ (indexed by $j$) items. Without loss of generality, we focus on the implicit user feedback which is more common in practical recommendation scenarios~\cite{qin2020sequential,wang2019kgat}. Particularly, in tensor $\mathcal{\boldmath{X}}$, each entry $x_{i,j,l}=1$ if user $u_i$ is interacted with item $t_j$ given the $l$-th behavior type. For example, if user $u_i$ purchases item $t_j$, the corresponding element $x_{i,j,l}$ will be set as 1 in the purchase behavior matrix $\mathcal{\boldmath{X}}_l$.
\end{Dfn}

\noindent \textbf{Problem Statement}. Based on the aforementioned definitions, the recommendation task with multiplex behavior learning is formulated as follows: \textbf{Input}: the user-item interaction data represented with multi-behavior tensor $\mathcal{\boldmath{X}}$ (including both the target and source behavior). \textbf{Output}: A predictive model to effectively infer the unknown user-item interactions in $\mathcal{\boldmath{X}}$ with the target behavior $l$.
\begin{align}
\text{Pr}(x_{i,j,l}=1) = f(\mathcal{\boldmath{X}} \in \mathbb{R}^{I\times J\times L})~~i\in [1,...,I];j\in [1,...,J]
\end{align}

\section{Methodology}
\label{sec:solution}

\begin{figure*}
    \centering
    \includegraphics[width=\textwidth]{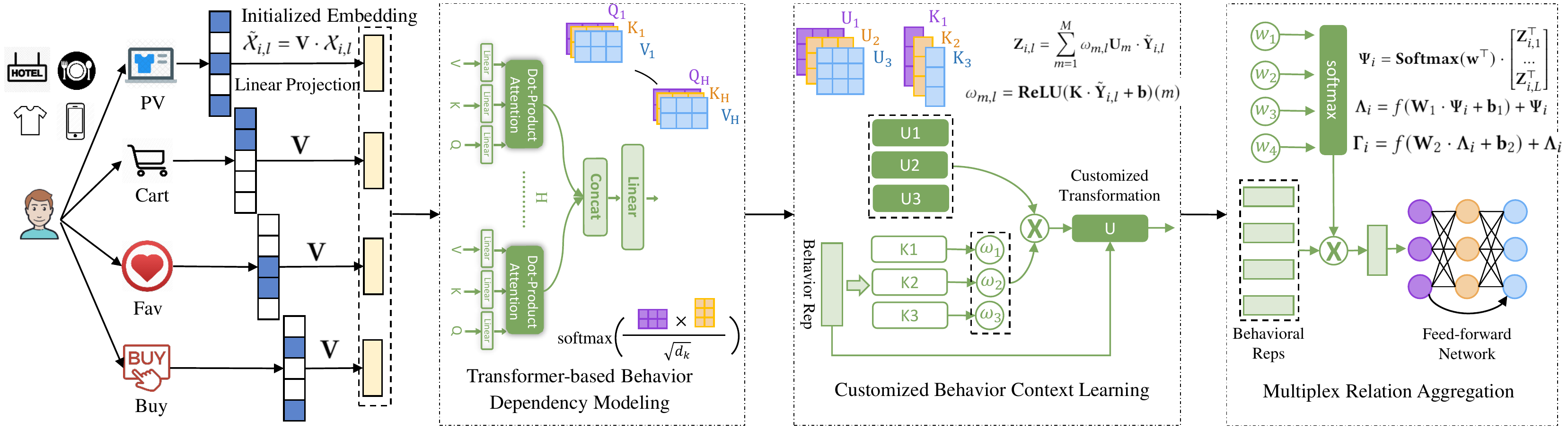}
    \caption{The model architecture of the proposed \model\ framework. The initialized embedding layer shares parameters across different behavior types. The transformer-based behavior dependency encoder takes all kinds of behavioral interaction data for dependency modeling. Different types of behaviors are individually transformed by the customized context learning with shared key and memory slots. $\bigotimes$ is the dot-product between the embeddings and transformation weight matrix.}
    \label{fig:framework}
\end{figure*}

In this section, we present the technical details of \model\ framework, the architecture of which is illustrated in Figure~\ref{fig:framework}. \model\ is a hierarchical neural architecture with three key modules in \model: (i) cross-behavior embedding layers that learn the representations by exploring the inter-dependencies across different types of interactions; (ii) a customized representation recalibration network that refines the latent embeddings, with the preservation of individual behavioral contextual information; (iii) a forecasting layer that aggregates the refined behavior type-specific embeddings and outputs a predicted likelihood of a user-item interaction pair.

\subsection{Multi-Behavior Dependency Modeling}
As discussed before, different types of user behaviors are correlated with each other, which brings in new challenges to the recommendation framework. To model the inter-dependencies across different types of behavior, we design a multi-behavior transformer network to promote the collaboration of different behavioral views. To achieve this goal, we learns a robust representation for user-item interactive patterns of each individual categorical behavior $l$, which integrates the relevant information from other behavior views $l'\in [1,...,L] \& l'\neq l$.

\subsubsection{\bf Initialized Embedding Layer}
Firstly, a projection layer is introduced to map the original multi-behavior user-item interaction data into initial latent representations. We denote the interaction vector of $l$-th behavior type and $i$-th user ($u_i$) over all items ($t_j,1\leq j \leq J$) as $\mathcal{\boldmath{X}}_{i,l} \in \mathbb{R}^J$. The projection operation for $\mathcal{\boldmath{X}}_{i,l}$ is formally defined as $\tilde{\mathcal{\boldmath{X}}}_{i,l} = \textbf{V} \cdot \mathcal{\boldmath{X}}_{i,l}$, where $\textbf{V}\in \mathbb{R}^{d\times J}$ and $d$ denotes learned projection matrix and hidden state dimensionality, respectively. Note that $\textbf{V}$ is shared across behavior categories to model the common semantics of different interactions. The projected $\tilde{\mathcal{\boldmath{X}}}_{i,l}$ serves as an initial parameterized state for user-item interactions $\mathcal{\boldmath{X}}_{i,l}$, to be optimized with the following modules.

\subsubsection{\bf Multi-Head Self-Attentive Mechanism}
Inspired by the promising potential of self-attention mechanism in data correlation learning~\cite{zhang2019next}, we build our multi-behavior dependency learning module upon the architecture of multi-head self-attention network, which allows the learned behavior type-specific representations to interact with each other and identify the most informative correlated signals across different types of interaction behavior. Furthermore, considering the fact that different types of interaction behavior (\eg, add-to-cart and purchase) can be mutually correlated in a complex way (due to personalized factors)~\cite{cen2019representation}, the multi-head learning strategy enable our behavior dependency encoder with the capability of jointly attending to information from different representation subspaces~\cite{yun2019sain}. In our transformer network, we adopts the scaled dot-product attention for each $h$-th head with the definitions of query, key and value transformation matrices $\textbf{Q}^h \in \mathbb{R}^{\frac{d}{H}\times d}$, $\textbf{K}^h \in \mathbb{R}^{\frac{d}{H}\times d}$ and $\textbf{V}^h \in \mathbb{R}^{\frac{d}{H}\times d}$. Then, the weight $\hat{\alpha}_{l,l'}^h$ assigned to each input value is determined by the dot-product of the query with all the keys as follows:
\begin{align}
\label{eq:att_w}
\alpha_{l,l'}^h=\frac{(\textbf{Q}^h \cdot \tilde{\mathcal{\boldmath{X}}}_{i,l})^\top (\textbf{K}^h \cdot \tilde{\mathcal{\boldmath{X}}}_{i,l'})}{\sqrt{\frac{d}{H}}};~~~\hat{\alpha}_{l,l'}^h = \frac{\exp{\alpha}^h_{l,l'}}{\sum_{l'=1}^L \exp{\hat{\alpha}^h_{l,l'}}}
\end{align}
\noindent where $\alpha_{l,l'}^h$ is the intermediate variable fed into the softmax operation to generate the final relevance score $\hat{\alpha}_{l,l'}^h$ between the $l$-th and $l'$-th type of behavior. Based on the learned head-specific attention weights, our dependency encoding module aims to learns a cross-head relevance score for each behavior type-specific representation $\tilde{\mathcal{\boldmath{X}}}_{i,l}$ with the following multi-head learning operations:
\begin{align}
\textbf{Y}_{i,l} = \text{MH-Att}(\tilde{\mathcal{\boldmath{X}}}_{i,l})=\mathop{\Bigm|\Bigm|}\limits_{h=1}^H \sum_{l'=1}^L  \alpha_{l,l'}^h\textbf{V}^h\cdot \tilde{\mathcal{\boldmath{X}}}_{i,l'}
\end{align}
To alleviate the gradient vanishing issue, the residual connection~\cite{he2016deep} is employed in the deep neural network structures. Additionally, we element-wisely add the learned dependency-aware behavior type-specific interaction representation $\textbf{Y}_{i,l}$ with the projected feature embedding $\tilde{\mathcal{\boldmath{X}}}_{i,l}$ of $l$-th behavior, so as to jointly preserve the behavior type-specific interaction features and the underlying inter-dependent signals across various types of user behavior. Formally, such operation is given as: $\tilde{\textbf{Y}}_{i,l} = \tilde{\mathcal{\boldmath{X}}}_{i,l} + \textbf{Y}_{i,l}$.

% and to preserve the original behavior type-specific features, the residual connection~\cite{he2016deep} is employed and the representation becomes: $\tilde{\textbf{Y}}_{i,l} = \tilde{\mathcal{\boldmath{X}}}_{i,l} + \textbf{Y}_{i,l}$.
% To capture the non-linear feature interactions, we further equip our multi-behavior transformer network with a two-layer feed-forward network which is jointly trained with other parts of the behavior dependency encoder. Formally, we give our feed-forward network structure as below:
% \begin{align}
%     \tilde{\textbf{Y}}_l=f(\textbf{W}_2\cdot f(\textbf{W}_1 \cdot(\tilde{\mathcal{\boldmath{X}}}_l+\textbf{Y}_l)+\textbf{b}_1) +\textbf{b}_2) + \tilde{\mathcal{\boldmath{X}}}_l+\textbf{Y}_l
% \end{align}
% where $\textbf{W}_1, \textbf{W}_2\in\mathbb{R}^{d\times d},\textbf{b}_1,\textbf{b}_2\in\mathbb{R}^d$ are transformations and bias vectors. In order to alleviate the gradient vanishing issue, the residual connection~\cite{he2016deep} is employed in our transformer network. 

% Then, we build upon the architecture of transformer network to encode the inter-dependencies across behavior categories.

\subsection{Customized Behavioral Context Learning}
In addition to the implicit multi-behavior dependency encoded by the above introduced transformer network, each type of behavior may have its own characteristics. For instance, users' page view behavior are more frequent than their purchases and add-to-cart behavior is more likely to be followed by a purchase than the add-to-favorite behavior. While the cross-behavior inter-correlation structure can be modeled by our transformer module, the behavior type-specific semantic diversity has been overlooked. Hence, we propose to augment our \model\ framework with the capability of capturing the semantic signals of each individual type of interaction behavior. Motivated by the recent advancements of augmented neural architecture and attention mechanism~\cite{tay2018latent,ma2019memory}, we perform a customized representation recalibration process on behavior type-specific context with a memory-augmented attention network. In our memory-based behavior context learning module, we provide a customized transformations for each type of user behavior representation $\tilde{\textbf{Y}}_{i,l}$ by stacking a set of memory blocks. By doing so, we endow \model\ with the power of distilling the underlying semantics from the specific contextual user-item interaction scenario (\eg, page view, interested in, want to buy, or purchase).

% The transformer-based behavioral embedding layers enhance the behavioral representations by modeling their inter-dependencies. However, the module is not capable enough to fully excavate the hidden information of multi-behavioral data, as the networks overlook the semantic diversity among different interaction types. Using its shared embedding and transformer for different behaviors, the dependency modeling module tends to learn a general model for all categories but probably not specialized enough. To this end, we equip \model\ with a memory-augmented attention~\cite{tay2018latent, xin2019relational} module which is capable of learning customized transformations for different inputs by attending on augmented memory blocks. The module is employed to learn behavior-type-specific contexts with different semantics and refine the behavioral representations.

In specific, our customized embedding recalibration module aims to learn $M$ (indexed by $m$) transformation matrices (individual is referred as $\textbf{U}_m\in\mathbb{R}^{d\times d}$) as the corresponding augmented memory, in order to project the general behavior embedding $\tilde{\textbf{Y}}_{i,l}$ into a type-aware latent learning space. By applying different memory transformations over different types of behavior, each type of behavioral features are refined with respect to its own contexts with the designed memory, and a customized behavioral representations are generated through this type-specific transformation procedure.

Furthermore, in order to alleviate the overfitting phenomenon of type-specific memory augmented neural network architecture~\cite{zheng2016neural}, we employ an attention network to learn the relations between $L$ behavior types and $M$ memory matrices in an explicit way, and generate a behavior type-specific transformations with weighted summation. Formally, the refined representation with customized behavioral context for the $l$-th interaction type is
\begin{align}
    \textbf{Z}_{i,l}=\sum_{m=1}^M\omega_{m,l}\textbf{U}_m\cdot\tilde{\textbf{Y}}_{i,l};~~~\omega_{m,l}=\textbf{ReLU}(\textbf{K}\cdot\tilde{\textbf{Y}}_{i,l}+\textbf{b})(m)
\end{align}
where $\textbf{K}\in\mathbb{R}^{M\times d},\textbf{b}\in\mathbb{R}^M$ are the transformation and bias for calculating attention weights. Instead of using Softmax, we use ReLU to relieve the gradient vanishing issue and mke it easier to train the attentive weight calculating. The memory transformation matrices $\textbf{U}_m$ and calculating attention weights $\omega_{m,l}$ are jointly trained with other components of \model.

\subsection{Multiplex Relation Aggregation Layer}
Next, we build upon a behavior type-wise gating mechanism to aggregate the learned latent representations from the memory-augmented transformer network, with the exploration of their contributions in capturing user's preference and assisting making predictions on the target behavior. Considering the distinct effects of different types of behavior in characterizing user's interest, \eg\ user's historical purchases may be more relevant to his future purchases as compared to his page view activities, the type-specific importance is learned in our gated mechanism in an adaptive manner. Formally, the applied weighted aggregation gate outputs a $d$-dimensional unified representation for $u_i$ as follows:
\begin{align}
    \mathbf{\Psi}_{i}= \textbf{Softmax}(\textbf{w}^\top)\cdot
    \begin{bmatrix} \textbf{Z}_{i,1}^\top \\ ... \\ \textbf{Z}_{i,L}^\top \end{bmatrix}
\end{align}
where $\textbf{w}\in\mathbb{R}^d$ is the parametric weights for aggregation, the Softmax activation function is used to normalize the weights. By applying the weighted aggregation gate, \model\ learns the contributions of different behavior types and thus can enable the adaptive aggregation in modeling the cross-type behavior relations.

% Having the refined behavioral representations $\textbf{Z}_{i,l}$, \model\ next summarizes the features using an aggregation gate. Considering the distinct effects of different types of behavior in characterizing user's interest, \eg\ user's historical purchases may be more relevant to his future purchases as compared to his page view activities, the type-specific importance is learned in our gated  adaptive weighted scheme. Formally, the applied weighted aggregation gate outputs a $d$-dimensional unified representation for $u_i$ as
% \begin{align}
%     \mathbf{\Psi}_{i}= \textbf{Softmax}(\textbf{w}^\top)\cdot
%     \begin{bmatrix} \textbf{Z}_{i,1}^\top \\ ... \\ \textbf{Z}_{i,L}^\top \end{bmatrix}
% \end{align}
% where $\textbf{w}\in\mathbb{R}^d$ is the parametric weights for aggregation, the Softmax activation function is used to normalize the weights. By applying the weighted aggregation gate, \model\ learns the contributions of different behavior types and can thus make adaptive aggregation.

After obtaining the aggregated user behavior representation, the \model\ adopts a two-layer feed-forward network with non-linear activation, to capture the complex feature interactions in the latent embeddings. Formally the deeply-extracted user representations are learned with the following operation:
\begin{align}
    \label{eq:dense}
    \mathbf{\Lambda}_i=f(\textbf{W}_1\cdot\mathbf{\Psi}_i+\textbf{b}_1)+\mathbf{\Psi}_i;~~
    \mathbf{\Gamma}_i=f(\textbf{W}_2\cdot\mathbf{\Lambda}_i+\textbf{b}_2)+\mathbf{\Lambda}_i
\end{align}
where $\textbf{W}_*\in\mathbb{R}^{d\times d}$ and $\textbf{b}_*\in\mathbb{R}^{d}$ are transformation and bias vectors of the neural network, $f$ is element-wisely applying non-linear activation functions, and residual connections are also employed. $\mathbf{\Gamma}_i\in\mathbb{R}^d$ is the final user representation.

\subsection{The Learning Process of \emph{\model}}
% With the user representation extracted from multiplex behavior data, \model\ makes prediction on one user $u_i$'s preferences over all items with target behavior $l$ through a linear projection $\text{Pr}(\mathcal{\boldmath{X}}_{i,l})=\textbf{P}\cdot\mathbf{\Gamma}_i$, where $\textbf{P}\in\mathbb{R}^{J\times d}$ is a parametric matrix composed of item embeddings and the result $\text{Pr}(\mathcal{\boldmath{X}}_{i,l})$ is a $J$-dimensional vector containing preference predictions for all items.

Given the user behavior representation aggregated from different views (\ie, behavior type-specific semantics and cross-type behavior dependencies), \model\ could make predictions on user's preference over items for the target $l$-th type of behavior. In particular, the prediction process is performed through a dot product operation $\text{Pr}(\mathcal{\boldmath{X}}_{i,j,l})=\textbf{P}_j^\top\cdot\mathbf{\Gamma}_i$, where $\textbf{P}_j\in\mathbb{R}^d$ is from a parametric embedding table for all items, and the result $\text{Pr}(\mathcal{\boldmath{X}}_{i,j,l})$ is a scalar score representing $u_i$'s tendency of interacting with $t_j$ under behavior $l$. Inspired by the settings of learning process on top-N recommendation tasks~\cite{zheng2019deep,nikolakopoulos2019recwalk}, we leverage the pair-wise loss to model the relative position in ranking-based recommendation scenarios. For each training step for user $u_i$, we sample a positive interaction set $\{t_{p_1},t_{p_2},...t_{p_s}\}$ composed of interacted items with $u_i$ for the target $l$-th type of behavior. Here. $s$ is defined as the number of the positive samples. Correspondingly, the same number of items that have no interactions with $u_i$ in the training set are randomly sampled to form the negative interaction set $\{t_{n_1},t_{n_2},...,t_{n_s}\}$. Based on the above descriptions, we formally define our loss function over all the samples of all users as below:
\begin{align}
    \text{Loss} = \sum_{i=1}^I\sum_{k=1}^s\textbf{max}(0,1-\text{Pr}(\mathcal{\boldmath{X}}_{i,p_k,l})+\text{Pr}(\mathcal{\boldmath{X}}_{i,n_k,l}))
    +\lambda\|\mathbf{\Theta}\|_\text{F}^2
\end{align}
\noindent where the first term is the pair-wise loss for a positive-negative pair. It expands the signed difference between two predictions, until it reaches a big enough scale. The latter term is a weight decay regularization term to prevent over-fitting, and $\lambda$ is the regularization weight. The learning process is elaborated in Algorithm~\ref{alg:learn_alg}.

\begin{algorithm}[t]
	\caption{Learning Process of \model\ Framework}
	\label{alg:learn_alg}
	\LinesNumbered
	\KwIn{user-item interaction tensor $\mathcal{\boldmath{X}}\in\mathbb{R}^{I\times J\times L}$, target behavior $l$, sample number $s$, maximum epoch number $E$, regularization weight $\lambda$, learning rate $\eta$}
	\KwOut{trained parameters in $\mathbf{\Theta}$}
	Initialize all parameters in $\mathbf{\Theta}$\\
    \For{$e=1$ to $E$}{
        Draw a mini-batch $\textbf{U}$ from all users $\{1,2,...,I\}$\\
        $\text{Loss} = \lambda\cdot\|\mathbf{\Theta}\|_\text{F}^2$\\
        \For{each $u_i\in\textbf{U}$}{
            Sample $s$ positive items $\{t_{p_1},...,t_{p_s}\}$ from $\mathcal{\boldmath{X}}_{i,l}$\\
            Sample $s$ negative items $\{t_{n_1},...,t_{n_s}\}$ from $\mathcal{\boldmath{X}}_i$\\
            Compute $\mathbf{\Gamma}_i$ according to Eq~\ref{eq:att_w} to Eq~\ref{eq:dense}\\
            \For{$k=1$ to $s$}{
                $\text{Pr}(\mathcal{ \boldmath{X}}_{i,p_k,l}) = \textbf{P}_{p_k}^\top \cdot \mathbf{ \Gamma }_i$\\
                $\text{Pr}(\mathcal{ \boldmath{X}}_{i,n_k,l}) = \textbf{P}_{n_k}^\top \cdot \mathbf{ \Gamma }_i$\\
            $\text{Loss}+=\textbf{max}(0, 1 - (\text{Pr}(\mathcal{\boldmath{X}}_{i,p_k,l}) - \text{Pr}(\mathcal{\boldmath{X}}_{i,n_k,l})))$\\
            }
        }
        \For{each parameter $\theta\in\mathbf{\Theta}$}{
            $\theta=\theta-\eta\cdot\partial\text{Loss}/\partial\theta$
        }
    }
    return all parameters $\mathbf{\Theta}$
\end{algorithm}

\section{Evaluation}
\label{sec:eval}
% In this section, we perform experiments on predicting two-levels of user preferences, often refered as \textit{click} and \textit{buy}. They corresponds to users' (vague) interests and (strong) purchase intentions, respectively.

In this section, we perform experiments on different datasets to demonstrate the effectiveness of our \emph{\model}. We aim to answer the following research questions:

%which lead the experiments and corresponding discussions:

\begin{itemize}[leftmargin=*]
\item \textbf{RQ1}: Compared to state-of-the-art models, does \emph{\model} achieve better performance in various recommendation applications?\\\vspace{-0.1in}
\item \textbf{RQ2}: What is the impact of the designed modules in \emph{\model}? Are the proposed cross-behavior transformer network and attention memory module necessary for improving performance?\\\vspace{-0.1in}
% \item \textbf{RQ3}: How does \emph{\model} perform with different settings of training/test data ratios, as compared to various baselines?
\item \textbf{RQ3}: How is the \emph{\model}'s recommendation accuracy \wrt\ the integration of different types of behavior?\\\vspace{-0.1in}
\item \textbf{RQ4}: What is the influence of hyperparameter settings in \emph{\model} for the recommendation performance?\\\vspace{-0.1in}
\item \textbf{RQ5}: What behavior relational patterns does the proposed \emph{\model} model capture for the final recommendation decision?\\\vspace{-0.1in}
\item \textbf{RQ6}: How is the scalability of the \emph{\model} framework?
\end{itemize}

\subsection{Experiment Settings}

\subsubsection{\bf Data Description}
\label{sec:data}
We evaluate the model performance on three different types of datasets: (i) MovieLens: a benchmark dataset for movie recommendations; (ii) Yelp: another benchmark dataset for location-based venue recommendations from the online review platform Yelp; (iii) E-Commerce: a user behavior data from a real-world e-commence platform. Table~\ref{tab:data} summarizes the data statistics and we present the data details as below:\\\vspace{-0.1in}

\begin{table}[t]
\vspace{-0.1in}
    \caption{Statistics of experimented datasets}
\vspace{-0.15in}
    \label{tab:data}
    \centering
    \footnotesize
	\setlength{\tabcolsep}{0.6mm}
    \begin{tabular}{ccccc}
        \toprule
        Dataset&User \#&Item \#&Interaction \#&Interactive Behavior Type\\
        \midrule
        Yelp&19800&22734&$1.4\times 10^6$&\{Tip, Dislike, Neutral, Like\}\\
        ML10M&67788&8704&$9.9\times 10^6$&\{Dislike, Neutral, Like\}\\
        E-Commerce&805506&584050&$6.4\times 10^7$&\{Page View, Favorite, Cart, Purchase\}\\
        \bottomrule
    \end{tabular}
\vspace{-0.1in}
\end{table}

% We evaluate the performance of the proposed \model\ on three large-scale real-world datasets,~\ie\ Yelp, MovieLens-10M (ML-10M) and UserBehavior. We filtered out users and items who have less than 5 observations or less than 2 \textit{buy} interactions. The statistics of resulted datasets are summarized in Table~\ref{tab:stat}. The datasets contain different scale of users, items and interactions, and have different behavior types, which can better validate the efficacy of the proposed method. Observed interactions of any kind are viewed as the \textit{click} behavior for all the datasets.

\noindent\textbf{MovieLens Data}\footnote{https://grouplens.org/datasets/movielens/10m/}.
It is a widely-used dataset for performance validation of various recommendation methods. Following the partition strategy in~\cite{ostuni2013top,lim2015top}, we differentiate the explicit user-item interactive behavior into three types in terms of user rating scores (\ie, ranging from 1 (worst) to 5 (best) stars with 0.5 star as increment): the original rating score $\leq 2$, $> 2$ and $< 4$, $\geq 4$ corresponds to the \textit{dislike}, \textit{neutral} and \textit{like} user behavior, respectively. In the MovieLens dataset, we regard the \emph{like} interaction as the target behavior and other interactions (\textit{dislike} and \textit{neutral}) as source behavior, because the positive interactions between users and items may be more useful for capturing user's preferences in recommendations~\cite{loni2019top}.\\\vspace{-0.1in}

% It is a widely-used dataset composed of 5-star ratings. To test \model\ on dealing with multi-type behaviors, we categorized ratings of 2-star or lower as \textit{dislike}, ratings of 4-star or higher as \textit{like}, and others as \textit{neutral}. \textit{Like} behavior is taken as the \textit{buy} behavior.

\noindent\textbf{Yelp Data}\footnote{https://www.yelp.com/dataset/download}.
This is another recommendation benchmark dataset collected from Yelp. We use the same multi-behavior differentiation strategy as the MovieLens data and partition the 5-star range rating behavior into \textit{dislike}, \textit{neutral} and \textit{like} user behavior. In addition to the user rating behavior, this data includes an additional tip behavior to indicate that user writes a tip on his/her visited venues. Similar to the MovieLens data, the target behavior in Yelp data is also set as the \emph{like} interaction and others are set as source behavior.\\\vspace{-0.1in}

% This is another popular dataset on recommendation. It contains \textit{dislike}, \textit{neutral} and \textit{like} behaviors made out of 5-star ratings similarly as ML-10M. There are also \textit{tip} relations indicating if the user wrote a tip about the item. \textit{Like} is used as the \text{buy} behavior.

\noindent \textbf{E-Commerce Data}. Besides the two benchmark datasets for movie and location-based venue recommendations, we also evaluate our \model\ framework in a real-world recommendation scenario with explicit multiple user behavior data from a major online retailing platform. Specifically, this data contains four types of interaction behavior, \ie, \textit{page view}, \textit{add-to-favorite}, \textit{add-to-cart} and \textit{purchase}. We consider the \textit{purchase} behavior as the target one, since the purchase is directly related with the conversion rate of recommendation in real-life E-commerce sites~\cite{guo2019buying}.

% \noindent\textbf{UserBehavior}\footnote{https://tianchi.aliyun.com/dataset/dataDetail?dataId=649}.
% This dataset was collected from TMall. It contains four types of interactions: \textit{page view} (PV) for users viewing items' detailed pages, \textit{favorite} for users labeling items as "favorite", \textit{cart} for adding items to one's shopping cart, and \textit{buy} for purchasing.

\subsubsection{\bf Evaluation Settings and Metrics}
In our experiments, we utilize the leave-one-out evaluation strategy which has been widely utilized in recommendation literature~\cite{he2016fast,he2017neuralncf}. Following their evaluation settings, we regard the latest interaction of each user as the test set and use the rest of data for training. For efficient and fair evaluation, we follow the common strategy in~\cite{tang2018personalized,kang2018self} to associate each ground truth item with 99 randomly sampled negative instances which have not interacted with the corresponding user.

We leverage two widely-used ranking metrics: \textit{Hit Ratio (HR@$k$)} and \textit{Normalized Discounted Cumulative Gain (NDCG@$k$)}~\cite{wang2019neural,chen2018sequential}, to investigate the ranking performance (top-$k$ ranked recommended items) of all compared methods. Note that higher HR and NDCG scores reflect better recommendation results. In our experiments, we also evaluate the model performance by varying the $k$ value.

\subsubsection{\bf Competitive Baselines}
\label{sec:baseline}
To perform a comprehensive performance validation, we compare our \emph{\model} with 12 baselines from six research lines, which are elaborated as follows:

\noindent \textbf{Conventional Matrix Factorization-based Recommendation}:
\begin{itemize}[leftmargin=*]
\item \textbf{BiasMF}~\cite{koren2009matrix}: This method is built upon the matrix factorization architecture with the incorporation of user and item biases.
\end{itemize}

\noindent \textbf{Neural Collaborative Filtering Models for Recommendation}:
\begin{itemize}[leftmargin=*]
\item \textbf{DMF}~\cite{xue2017deep}: It is a deep matrix factorization model which takes both the explicit and implicit feedback as the input.
\item \textbf{NCF}~\cite{he2017neural}: NCF aims to supercharge collaborative filtering with non-linear neural networks. We consider three variants of NCF \wrt\ user-item interaction encoders: \ie, Multilayer perceptron (\ie, NCF-M), concatenated element-wise-product branch (\ie, NCF-N) and the fixed element-wise product (\ie, NCF-G).
\end{itemize}

\noindent \textbf{Collaborative Filtering with Auto-Encoder}:
\begin{itemize}[leftmargin=*]
\item \textbf{AutoRec}~\cite{sedhain2015autorec}: It leverages a three-layer autoencoder to map user-item interactions into latent representations.
\item \textbf{CDAE}~\cite{wu2016collaborative}: In this autoencoder CF, an adaptive loss is incorporated into the embedding projection process for users/items.
\end{itemize}

\noindent \textbf{Neural Auto-regressive Recommendation Models}:
\begin{itemize}[leftmargin=*]
\item \textbf{CF-NADE}~\cite{zheng2016neural}: It enhances the autoregressive collaborative filtering with the parameter sharing between different ratings.
\item \textbf{CF-UIcA}~\cite{du2018collaborative}: It is a neural co-autoregressive framework to consider the structural correlation for both users and items.
\end{itemize}

\noindent \textbf{Graph Neural Network Recommendation Models}:
\begin{itemize}[leftmargin=*]
\item \textbf{ST-GCN}~\cite{zhang2019star}: It stacks encoder-decoder blocks using graph convolutional networks to learn embeddings of users and items.
\item \textbf{NGCF}~\cite{wang2019neural}: This approach explore the structural knowledge with the message-passing mechanism to capture the high-order connections in the user-item interaction graph.
\end{itemize}

\noindent \textbf{Recommendation with Multi-Behavior Learning}:
\begin{itemize}[leftmargin=*]
\item \textbf{NMTR}~\cite{gao2019neural}: It is a multi-task recommendation model which considers the behavior correlations in a cascaded manner.
\item \textbf{DIPN}~\cite{guo2019buying}: This model utilizes bi-directional recurrent network and attention mechanism to consider the correlations between the buying or browsing activities.
\end{itemize}

\subsubsection{\bf Parameter Settings}
In the latent learning space of \emph{\model} framework, we set the hidden state dimensionality $d$ as 16. In the multi-behavior transformer module, we set the number of attention heads for multi-dimensional learning as 2. Furthermore, the number of memory transformations is set as 8 in our customized behavior-specific context encoding. We implement our \emph{\model} with TensorFlow and use Adam optimizer for model optimization with the learning rate and batch size of $1e^{-3}$ and 32, respectively. The decay rate of 0.96 is applied for each epoch during the training phase. To reduce the overfitting effect, we adopt set weight decay as the regularization strategy with the selection from \{0.05, 0.01, 0.005, 0.001, 0\}. The depth of our feature extraction module is set as 3. For the baselines (\ie, NCF and NMTR) which employ the point-wise loss, we set the sampling ratio for positive and negative instances from the range of $1:1$ to $1:4$.

\begin{table}
\vspace{-0.1in}
	\caption{Prediction performance on Yelp (like behavior), MovieLens (like behavior) and E-Commerce (purchase behavior) data, in terms of \textit{HR@$k$} and \textit{NDCG@$k$} ($k=10$).}
\vspace{-0.1in}
	\centering
% 	\scriptsize
    \ssmall
    % \footnotesize
    % \small
	\setlength{\tabcolsep}{0.6mm}
	\begin{tabular}{|c|c|c|c|c|c|c|c|c|c|c|c|c|}
		\hline
		\multirow{3}{*}{Model}&
		\multicolumn{4}{c|}{Yelp Data}&\multicolumn{4}{c|}{MovieLens}&\multicolumn{4}{c|}{E-Commerce}\\
		\cline{2-13}
% 		&\multicolumn{2}{c|}{Buy}&\multicolumn{2}{c|}{Buy}&\multicolumn{2}{c|}{Buy}\\
% 		\cline{2-13}
		&HR&Imp&NDCG&Imp&HR&Imp&NDCG&Imp&HR&Imp&NDCG&Imp\\
		\hline
		\hline
		BiasMF&0.755&9.4\%&0.481&10.2\%&0.767&10.4\%&0.490&16.1\%&0.262&35.1\%&0.153&36.6\%\\
		\hline
		DMF&0.756&9.3\%&0.485&9.3\%&0.779&8.7\%&0.485&17.3\%&0.305&16.1\%&0.189&10.6\%\\
		\hline
		NCF-M&0.714&15.7\%&0.429&23.5\%&0.757&11.9\%&0.471&20.8\%&0.319&11.0\%&0.191&9.4\%\\
		\hline
		NCF-G&0.755&9.4\%&0.487&8.8\%&0.787&7.6\%&0.502&13.3\%&0.290&22.1\%&0.167&15.1\%\\
		\hline
		NCF-N&0.771&7.1\%&0.500&6.0\%&0.801&5.7\%&0.518&9.8\%&0.325&8.9\%&0.201&4.0\%\\
		\hline
		AutoRec&0.765&8.0\%&0.472&12.3\%&0.658&28.7\%&0.392&45.2\%&0.313&13.1\%&0.190&10.0\%\\
		\hline
		CDAE&0.750&10.1\%&0.462&14.7\%&0.659&28.5\%&0.392&45.2\%&0.329&7.6\%&0.196&6.6\%\\
		\hline
		CF-NADE&0.792&4.3\%&0.499&6.2\%&0.761&11.3\%&0.486&17.1\%&0.317&11.7\%&0.191&9.4\%\\
		\hline
		CF-UIcA&0.750&10.1\%&0.469&13.0\%&0.778&8.9\%&0.491&15.9\%&0.332&6.6\%&0.198&5.6\%\\
		\hline
		ST-GCN&0.775&6.6\%&0.465&14.0\%&0.738&14.8\%&0.444&28.2\%&0.347&2.0\%&0.206&1.5\%\\
		\hline
		NGCF&0.789&4.7\%&0.500&6.0\%&0.790&7.2\%&0.508&12.0\%&0.302&17.2\%&0.185&13.0\%\\
		\hline
		NMTR&0.790&4.6\%&0.478&10.9\%&0.808&4.8\%&0.531&7.2\%&0.332&6.6\%&0.179&16.8\%\\
		\hline
		DIPN&0.791&4.4\%&0.500&6.0\%&0.811&4.4\%&0.540&5.4\%&0.317&11.7\%&0.178&17.4\%\\
		\hline
		\emph{\model}&\textbf{0.826}& -- &\textbf{0.530}& -- &\textbf{0.847}& -- &\textbf{0.569}&--&\textbf{0.354}&--&\textbf{0.209}&--\\
		\hline
	\end{tabular}
	\label{tab:target_behavior_results}
	\vspace{-0.1in}
\end{table}

\subsection{Performance Comparison (RQ1)}

\subsubsection{\bf Performance on Target Behavior}
In the evaluation, we first perform experiments to separately make recommendations on venue, movie and online retailing products with three types of datasets and the results are shown in Table~\ref{tab:target_behavior_results} (``Imp'' indicates the relatively improvement ratio). We observe the remarkable performance improvement achieved by our \emph{\model} in predicting different types of behaviors. We attribute such improvements to exploration of the cross-type behavior dependencies which are neglected by most existing methods, although they attempt to model complex user-item interactive relations with various deep neural encoders (\eg, autoencoder, graph neural network, attention mechanism).

Additionally, by jointly analyzing the results among the three datasets, we find that the improvement of \emph{\model} on the E-Commerce data is the most significant with the largest data scale. This may be caused by the behavior diversity: the multiple behaviors from the E-Commerce site are constructed with four different types of behavior which may show strong ordinal relations between the target (purchase) and source behaviors (\eg, page view $\rightarrow$ add-to-cart $\rightarrow$ purchase) in the real-world online retailing systems. The consistent improvement across datasets with different user-item interaction densities, suggests the robustness of \emph{\model} in accurately learning user preference under different sparsity degrees.

Lastly, it is worth mentioning that although the correlations between behavior has been considered in recent recommendation solutions (\ie, NMTR and DIPN), they merely model the singular dimensional cascading correlations between multi-type interactions, and cannot comprehensively capture the arbitrary dependencies between different types of interaction with different items. Therefore, such oversimplification on the behavior dependency leads to suboptimal recommendation results.

\begin{table*}[h]
\vspace{-0.1in}
\caption{Overall recommendation performance in forecasting click behavior in terms of \textit{HR@$k$} and \textit{NDCG@$k$} ($k=10$).}
\vspace{-0.05in}
\centering
%\footnotesize
\small
\setlength{\tabcolsep}{1mm}
\begin{tabular}{|c|c|c|c|c|c|c|c|c|c|c|c|c|c|c|c|}
\hline
Data & Metric & BiasMF & ~DMF~ & NCF-M & NCF-G & NCF-N & AutoRec & ~CDAE~ & CF-NADE & CF-UIcA & ST-GCN & NGCF & NMTR & DIPN & \emph{\model}\\
\hline
\multirow{2}{*}{Yelp}
&HR & 0.809 & 0.801 & 0.770 & 0.808 & 0.812 & 0.745 & 0.753 & 0.771 & 0.808 & 0.796 & 0.810 & 0.794 & 0.816 & \textbf{0.848}\\
\cline{2-16}
&NDCG & 0.513 & 0.503 & 0.464 & 0.519 & 0.523 & 0.456 & 0.456 & 0.478 & 0.512 & 0.483 & 0.521 & 0.473 & 0.514 & \textbf{0.548}\\
\hline
\multirow{2}{*}{MovieLens}
&HR & 0.727 & 0.730 & 0.693 & 0.745 & 0.748 & 0.612 & 0.613 & 0.677 & 0.718 & 0.688 & 0.735 & 0.773 & 0.776 & \textbf{0.808}\\
\cline{2-16}
&NDCG & 0.456 & 0.461 & 0.427 & 0.470 & 0.473 & 0.361 & 0.360 & 0.421 & 0.442 & 0.425 & 0.468 & 0.497 & 0.499 & \textbf{0.535}\\
\hline
\multirow{2}{*}{E-Commerce}
&HR & 0.383 & 0.399 & 0.401 & 0.400 & 0.409 & 0.423 & 0.427 & 0.486 & 0.428 & 0.452 & 0.470 & 0.409 & 0.405 & \textbf{0.535}\\
\cline{2-16}
&NDCG & 0.228 &0.245 & 0.239 & 0.240 & 0.244 & 0.257 & 0.262 & 0.293 & 0.257 & 0.257 & 0.281 & 0.236 & 0.237 & \textbf{0.326}\\
        %  \hline
        %  \multirow{2}{*}{MovieLens}
        %  &HR&0.745&0.795&0.797&0.793&0.809\\
        %  \cline{2-7}
        %  &NDCG&0.460&0.505&0.510&0.501&0.513\\
        %  \hline
        %  \multirow{2}{*}{E-Commerce}
        %  &HR&0.745&0.795&0.797&0.793&0.809\\
        %  \cline{2-7}
        %  &NDCG&0.460&0.505&0.510&0.501&0.513\\
\hline
\end{tabular}
\vspace{-0.05in}
\label{tab:click_behavior}
\end{table*}

\subsubsection{\bf Overall Prediction Click Behavior}
We also conduct experiments to evaluate the recommendation performance of all compared methods by forecasting the overall user-item interaction (\ie, click behavior), since the accurate predictions on overall interactive behavior (\eg, including all page view, add-to-cart and purchase behavior) could also provide useful insights for recommendation scenarios which focus on optimizing the click rate. As shown in Table~\ref{tab:click_behavior}, our \emph{\model} still achieves the best performance on all datasets as compared to various types of baselines. This validation shows the potential of the overall prediction performance of \emph{\model} by jointly considering multi-type behavior of users.

\subsubsection{\bf Ranking Performance v.s. Top-$K$ Value}
We also evaluate the model ranking performance by varying the $K$ value in terms of HR@$K$ and NDCG@$K$. We compare \emph{\model} with the best performed method of each baseline categories (see Section~\ref{sec:baseline} for baseline description), and report the results of predicting the click and like behavior on Yelp data in Table~\ref{tab:vary_k}. We can observe that \emph{\model} consistently outperforms other representative baselines with different settings of $K$.

\begin{table}[h]
    \vspace{-0.05in}
	\caption{Ranking performance evaluation on Yelp dataset with varying Top-\textit{K} value in terms of \textit{HR@K} and \textit{NDCG@K}}
% 	\vspace{-0.10in}
	\centering
    % \ssmall
    \footnotesize
    % \small
	\setlength{\tabcolsep}{0.6mm}
	\begin{tabular}{|c|c|c|c|c|c|c|c|c|c|c|c|}
		\hline
		\multirow{2}{*}{Model}&\multirow{2}{*}{Metric}&\multicolumn{5}{c|}{Click}&\multicolumn{5}{c|}{Like}\\
		\cline{3-12}
		&&@1&@3&@5&@7&@9&@1&@3&@5&@7&@9\\
		\hline
		\hline
		\multirow{2}{*}{BiasMF}&HR
		&0.261&0.519&0.644&0.728&0.799&0.287&0.474&0.626&0.714&0.741\\
		\cline{2-12}
		&NDCG&0.261&0.409&0.460&0.490&0.508&0.287&0.378&0.432&0.461&0.474\\
		\hline
        \multirow{2}{*}{NCF-N}&HR
        &0.278&0.535&0.661&0.747&0.800&0.260&0.481&0.604&0.695&0.742\\
        \cline{2-12}
        &NDCG&0.278&0.426&0.474&0.505&0.519&0.260&0.396&0.444&0.477&0.492\\
        \hline
        \multirow{2}{*}{AutoRec}&HR
        &0.227&0.443&0.568&0.650&0.715&0.228&0.455&0.586&0.684&0.732\\
        \cline{2-12}
        &NDCG&0.227&0.343&0.398&0.426&0.440&0.228&0.362&0.410&0.449&0.462\\
        \hline
        \multirow{2}{*}{CF-NADE}&HR
        &0.236&0.466&0.597&0.682&0.744&0.265&0.508&0.642&0.720&0.784\\
        \cline{2-12}
        &NDCG&0.236&0.368&0.423&0.452&0.468&0.265&0.402&0.454&0.478&0.497\\
        \hline
        \multirow{2}{*}{CF-UIcA}&HR
        &0.261&0.501&0.640&0.723&0.784&0.235&0.449&0.576&0.659&0.731\\
        \cline{2-12}
        &NDCG&0.261&0.390&0.447&0.478&0.500&0.235&0.360&0.412&0.440&0.463\\
        \hline
        \multirow{2}{*}{ST-GCN}&HR
        &0.231&0.474&0.614&0.704&0.766&0.216&0.445&0.580&0.669&0.744\\
        \cline{2-12}
        &NDCG&0.231&0.369&0.426&0.458&0.478&0.216&0.347&0.400&0.430&0.454\\
        \hline
        \multirow{2}{*}{NMTR}&HR
        &0.203&0.459&0.608&0.700&0.767&0.214&0.466&0.610&0.700&0.762\\
        \cline{2-12}
        &NDCG&0.203&0.352&0.412&0.445&0.465&0.214&0.360&0.419&0.450&0.469\\
        \hline
        % \multirow{2}{*}{NGCF}&HR
        % &0.281&0.536&0.657&0.749&0.799&0.221&0.495&0.639&0.718&0.785\\
        % \cline{2-12}
        % &NGCF&0.281&0.427&0.476&0.510&0.526&0.221&0.398&0.457&0.488&0.501\\
        % \hline
        \multirow{2}{*}{\emph{\model}}&HR
        &\textbf{0.296}&\textbf{0.560}&\textbf{0.693}&\textbf{0.771}&\textbf{0.828}&\textbf{0.279}&\textbf{0.529}&\textbf{0.659}&\textbf{0.741}&\textbf{0.798}\\
        \cline{2-12}
        &NDCG&\textbf{0.296}&\textbf{0.447}&\textbf{0.500}&\textbf{0.529}&\textbf{0.545}&\textbf{0.279}&\textbf{0.423}&\textbf{0.477}&\textbf{0.507}&\textbf{0.524}\\
		\hline
	\end{tabular}
	\label{tab:vary_k}
	\vspace{-0.1in}
\end{table}

\begin{figure*}[t!]
    \centering
    \vspace{-0.1 in}
    \subfigure[][Yelp-HR@10]{
        \centering
        \includegraphics[width=0.15\textwidth]{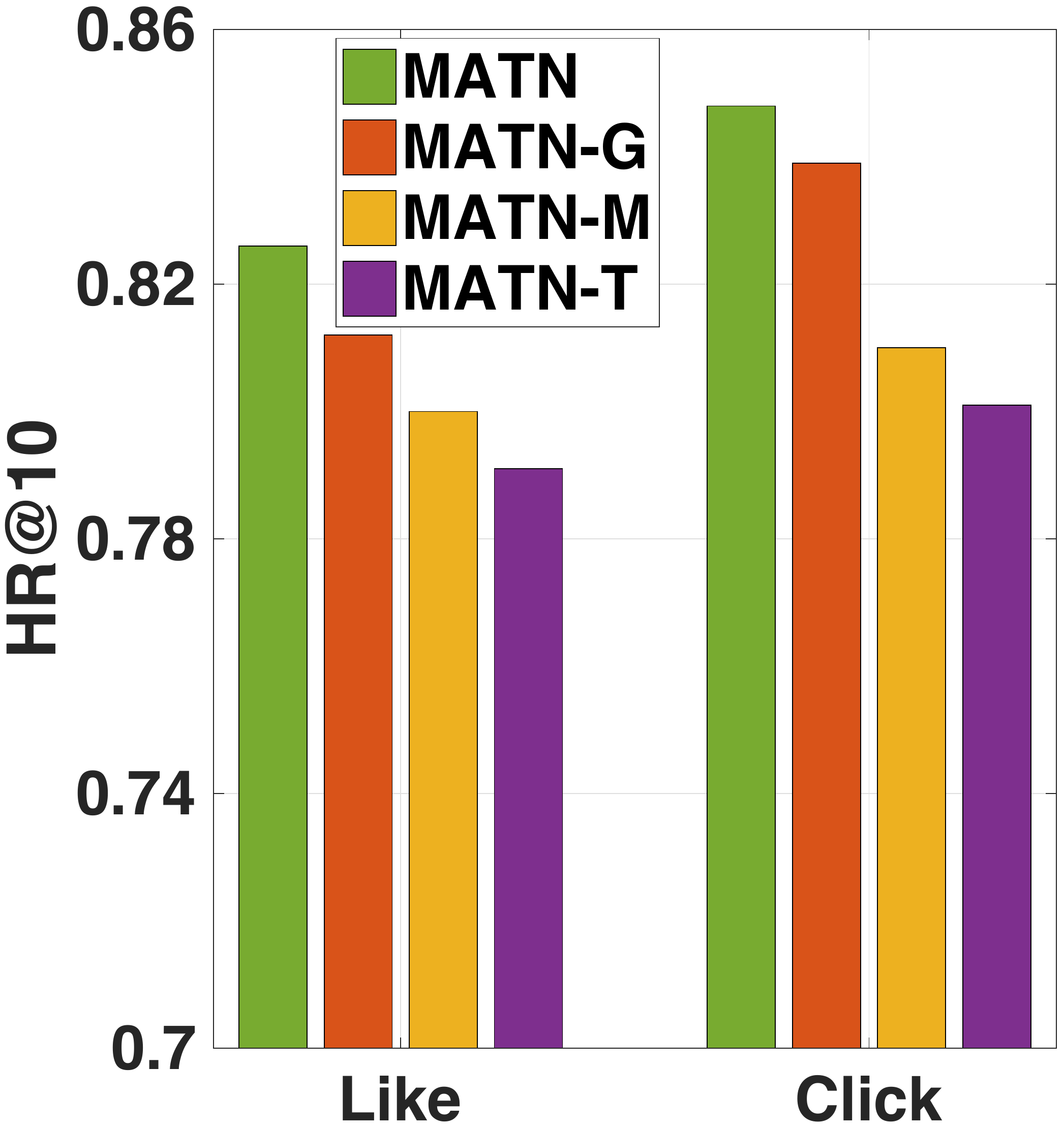}
        \label{fig:Yelp_HR}
        }
    \subfigure[][Yelp-NDCG@10]{
        \centering
        \includegraphics[width=0.15\textwidth]{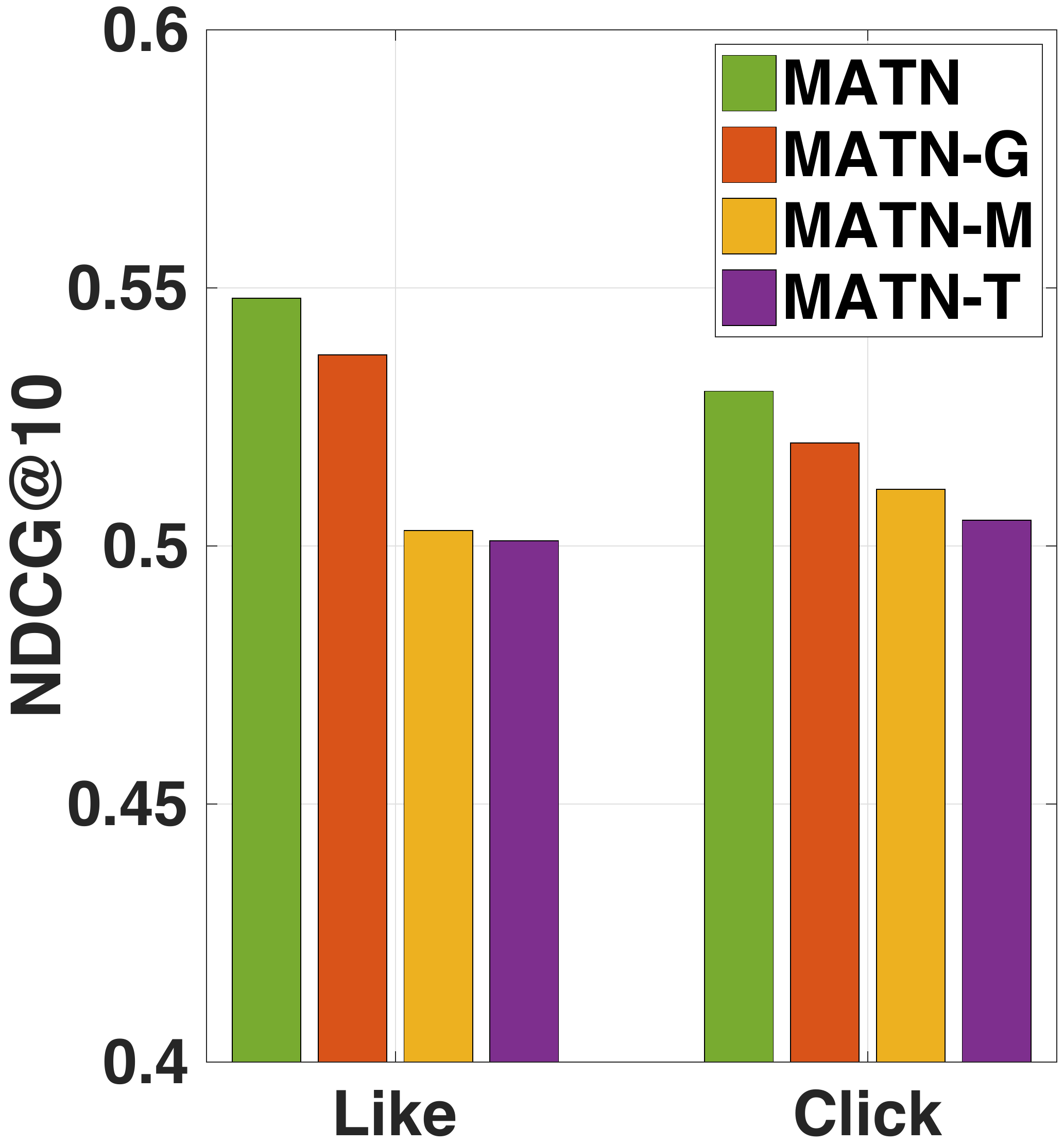}
        \label{fig:Yelp_NDCG}
        }
    \subfigure[][MovieLens-HR@10]{
        \centering
        \includegraphics[width=0.15\textwidth]{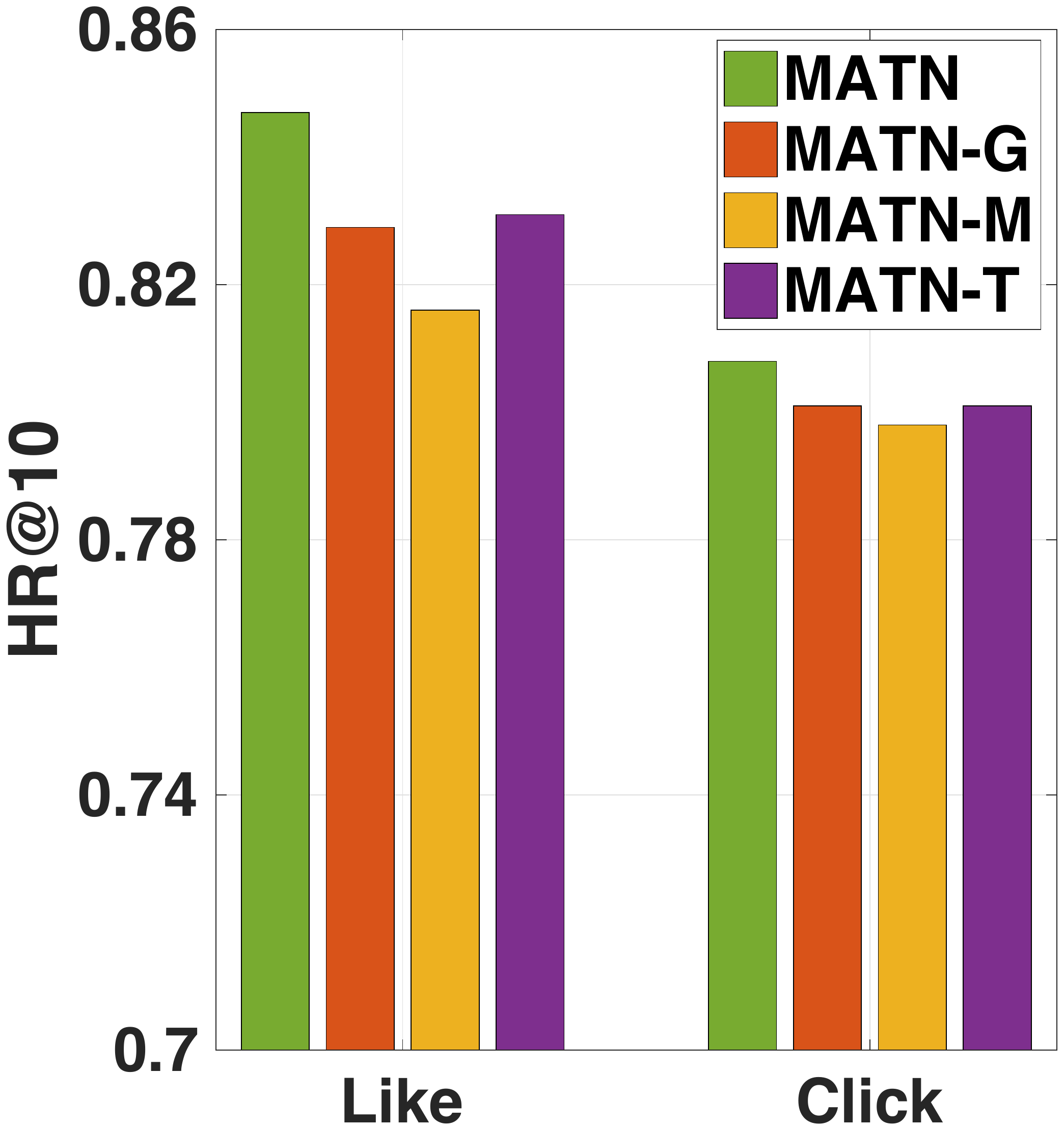}
        \label{fig:MovieLens_HR}
        }
    \subfigure[][MovieLens-NDCG@10]{
        \centering
        \includegraphics[width=0.15\textwidth]{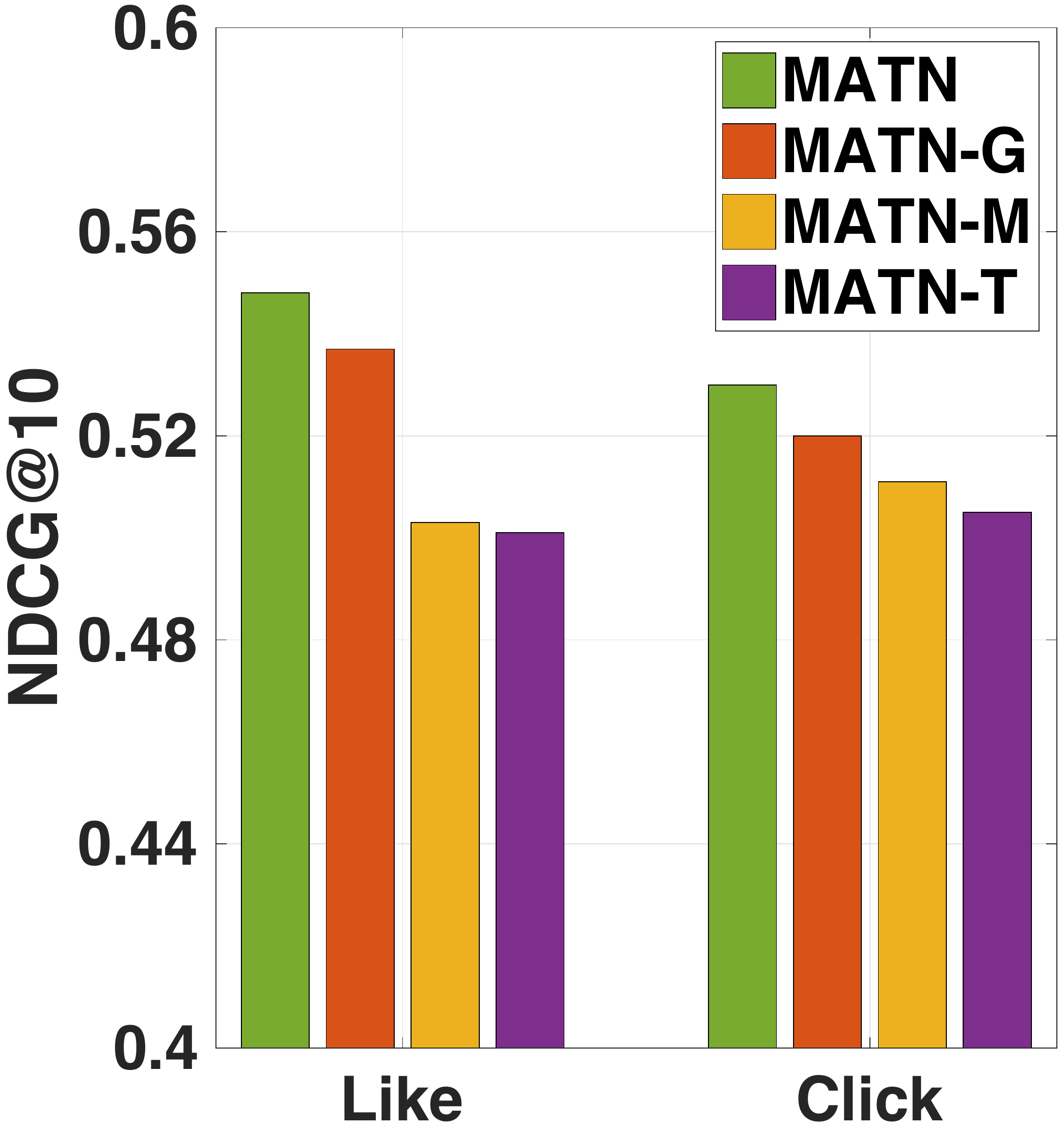}
        \label{fig:MovieLens_NDCG}
        }
    \subfigure[][E-Commerce-HR@10]{
        \centering
        \includegraphics[width=0.15\textwidth]{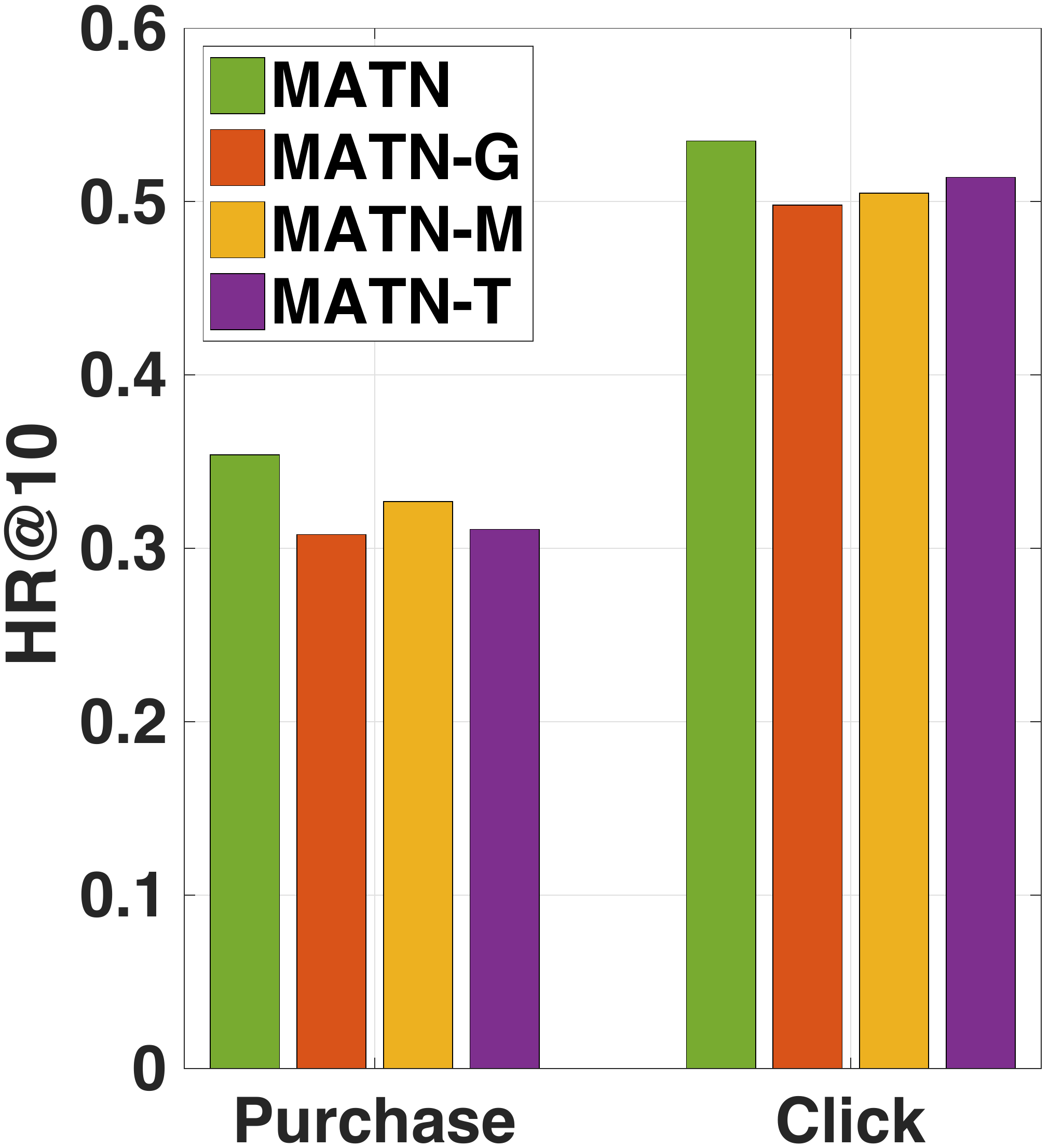}
        \label{fig:E_Commerce_HR}
        }
    \subfigure[][E-Commerce-NDCG@10]{
        \centering
        \includegraphics[width=0.15\textwidth]{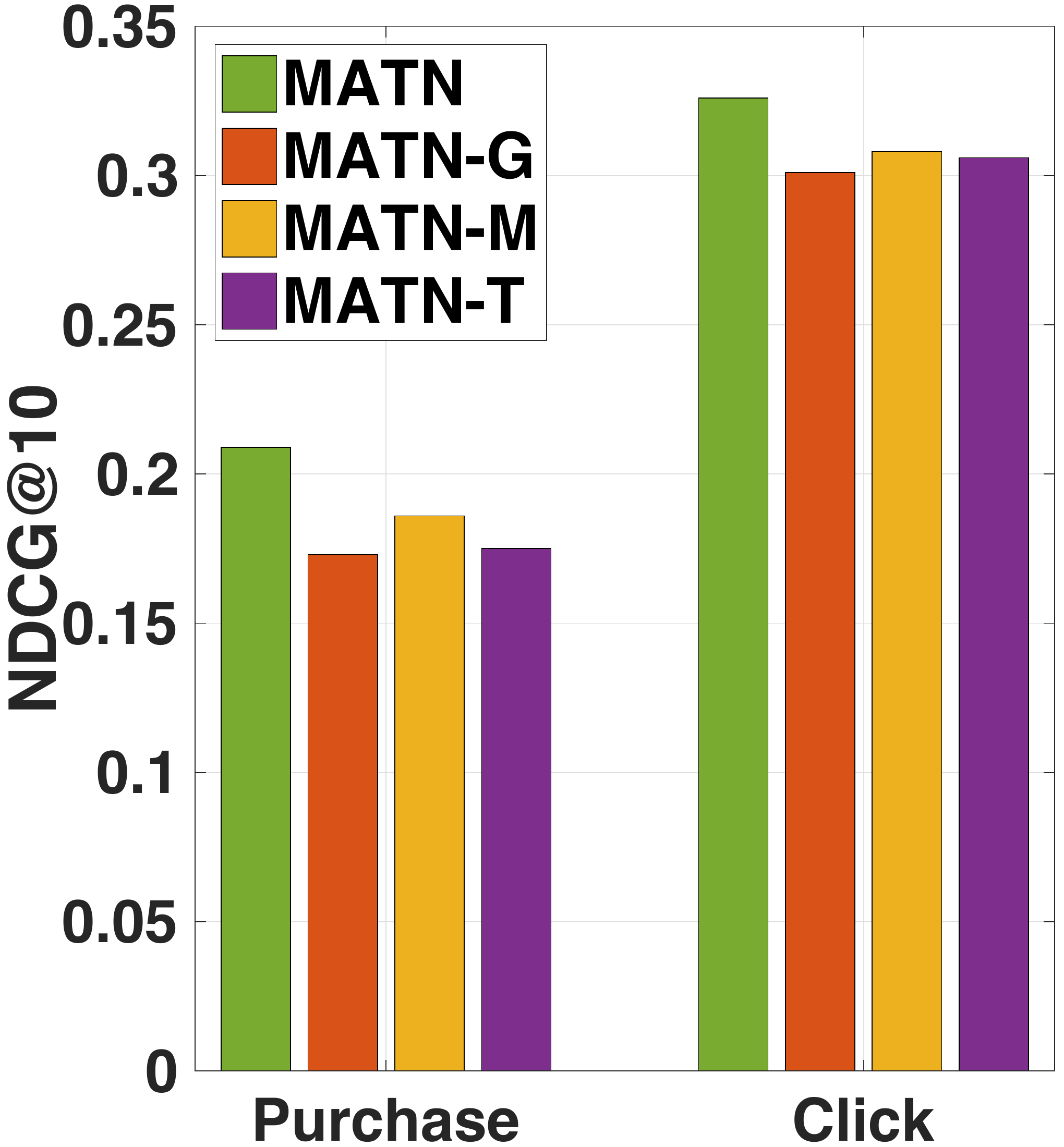}
        \label{fig:E_Commerce_NDCG}
        }
    \vspace{-0.1in}
    \caption{Model ablation study of \emph{\model} on Yelp, MovieLens and E-Commerce data in terms of HR@$K$ and NDCG@$K$.}
    \label{fig:variant_M}
    \vspace{-0.15in}
\end{figure*}

% \begin{figure}
%     \centering
%         \begin{adjustbox}{max width=\linewidth}
%             \input{./figures/ranking_performance_vary_k}
%         \end{adjustbox}
%         % \begin{adjustbox}{max width=\linewidth}
%         %     \input{./fig/ranking_metric_map_2}
%         % \end{adjustbox}
%         \vspace{-0.25in}
%         \caption{Ranking performance comparison in terms of \emph{HR@$k$} and \emph{NDCG@$k$}.}
%         \vspace{-0.15in}
%         \label{fig:pred_map}
% \end{figure}

\subsection{Model Ablation Study (RQ2)}
Furthermore, we conduct ablation experiments over a several key components of \emph{\model} to better understand the component-specific effects. Particularly, we introduce the following model variants:
\begin{itemize}[leftmargin=*]
\item \textbf{Effect of Multi-Behavior Transformer Network}: \emph{\model}-T. We do not utilize the multi-behavior transformer network to capture mutual relations between different types of behavior.
\item \textbf{Effect of Memory Attention Mechanism}: \emph{\model}-M. We remove the memory-augmented attention network in the joint \emph{\model} model to encode behavior type-specific semantics. 
\item \textbf{Effect of Gating Mechanism}: \emph{\model}-G. We replace the designed gating mechanism with the simplified average pooling operation over all behavior type-specific representations in the behavioral pattern aggregation layer.
\end{itemize}

Figure~\ref{fig:variant_M} presents the model ablation study results. We summary the following findings (\emph{\model} is the default model version). \\\vspace{-0.1in}

\noindent (1) The incorporation of mutual dependencies between different types of interaction behavior over all items, is capable of boosting the performance substantially. It demonstrates the rationality of our multi-head self-attention architecture in learning explicit pair-wise relations between different behavior types. \\\vspace{-0.1in}

\noindent (2) \emph{\model} is consistently superior to \emph{\model}-M, which hence illustrates the importance of considering context and semantics of individual type of behavior in profiling user preferences. \\\vspace{-0.1in}

\noindent (3) The replacement of our gating mechanism (\emph{\model}) with average pooling operation (\emph{\model}-G), degrades the model's performance. It make sense since \emph{\model}-G fails to model the different importance across different types of behavior in making final recommendations.

\subsection{Impact Studies of Multi-Behavior Relation Integration (RQ3)}
To investigate whether exploiting multi-type interaction behavior helps to achieve better performance, we further perform ablation experiments for the purchase prediction task on E-Commerce data, to show the effect of incorporating different types of user-item interactions in our \emph{\model} with four model variants: \emph{\model}$_F$--without the \emph{add-to-favorite} behavior; \emph{\model}$_C$--without the \emph{add-to-cart} behavior; and \emph{\model}$_P$--without the \emph{page view} behavior. Furthermore, we design another variant by removing all other types of interactions and only contain the purchase behavior \emph{\model}$_B$.

\begin{figure}
    \centering
    \subfigure[HR@\textit{k}]{
        \centering
        \includegraphics[width=0.475\columnwidth]{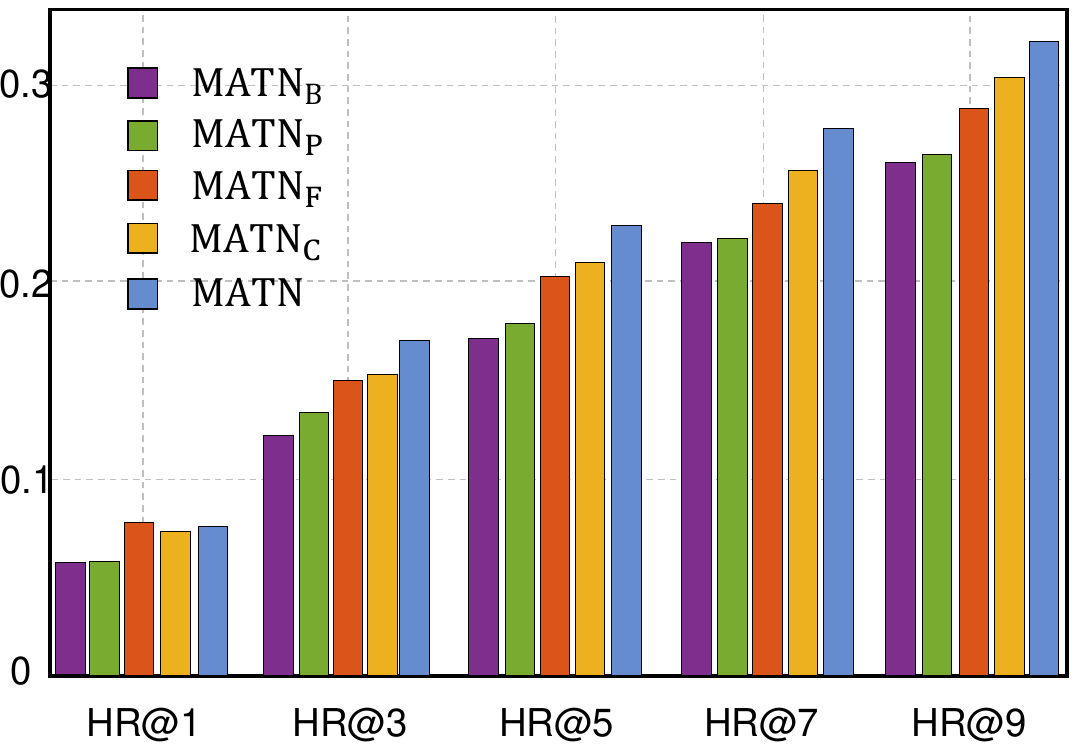}
    }
    \subfigure[NDCG@\textit{k}]{
        \centering
        \includegraphics[width=0.475\columnwidth]{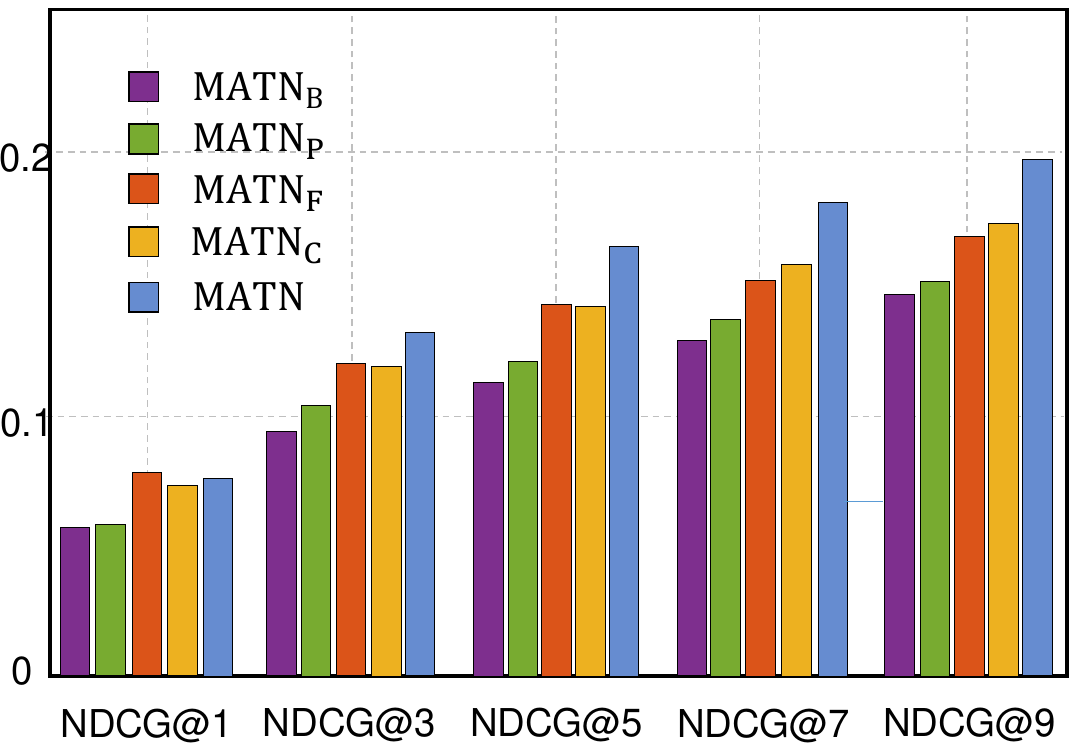}
    }
    \vspace{-0.15in}
    \caption{Impact study of multi-behavior relation integration on purchase prediction of E-Commerce dataset.}
    \vspace{-0.1in}
    \label{fig:behavior_combine}
\end{figure}

Figure~\ref{fig:behavior_combine} shows the evaluation results of different variants under varying top-k settings. We summarize the following findings:

\noindent (1) \emph{\model} using all types of interaction behaviors consistently outperforms other variants with varying top-\textit{k} settings, except for one exception on top-1 prediction with minor performance defect. The results validate that our \emph{\model} improve purchase forecasting through integrating multi-behavior relations.

\noindent (2) \emph{\model}$_B$ using only purchase data yields worst performance, which shows the positive contribution of the three additional behavior types (\ie\ page view, add-to-favorite and add-to-cart) in helping with user modeling in the e-commerce scenario.

\noindent (3) Among the three variants that utilize two additional behavior types (\ie\ \emph{\model}$_F$, \emph{\model}$_C$, \emph{\model}$_P$), \emph{\model}$_P$ clearly shows more severe performance degradation compared to the other two. This sheds light on the higher importance and effectiveness of utilizing page view data in \emph{\model} and online shopping recommendation.

\subsection{Hyperparameter Study of \emph{\model} (RQ4)}
\begin{figure*}
\vspace{-0.1in}
    \centering
    \begin{adjustbox}{max width=1.0\linewidth}
    \begin{tikzpicture}
\begin{axis}[
    xlabel={Hidden State Dimensionality $d$},
    ylabel={HR},
    xmin=4,xmax=32,
    ymin=0.700,ymax=0.870,
    legend columns=1,
    legend cell align=right,
    grid=both,
    every axis plot/.append style={ultra thick},
    every tick label/.append style={scale=1.3},
    label style={scale=2},
    legend style={
        nodes={scale=1.5, transform shape},
        legend image post style={scale=1.5},
        },
    legend style={at={(0,0)},anchor=south west},
    every axis plot post/.append style={
        every mark/.append style={scale=2}
    }
]

\addplot[color={rgb:red,4;green,2;yellow,1}, mark=otimes, dashed]
table[x=para, y=hr_click] {latFactor_click.txt};
\addplot[color={rgb:blue,4;green,2;yellow,1}, mark=o, dashed, mark options={solid}]
table[x=para, y=hr_buy] {latFactor_click.txt};
\legend{\Large Click, \Large Like};

\end{axis}
\end{tikzpicture}

\begin{tikzpicture}
\begin{axis}[
    xlabel={Hidden State Dimensionality $d$},
    ylabel={NDCG},
    xmin=4,xmax=32,
    ymin=0.400,ymax=0.580,
    legend columns=1,
    legend cell align=right,
    grid=both,
    every axis plot/.append style={ultra thick},
    every tick label/.append style={scale=1.3},
    label style={scale=2},
    legend style={
        nodes={scale=1.5, transform shape},
        legend image post style={scale=1.5},
        },
    legend style={at={(0,0)},anchor=south west},
    every axis plot post/.append style={
        every mark/.append style={scale=2}
    }
]

\addplot[color={rgb:red,4;green,2;yellow,1}, mark=otimes, dashed]
table[x=para, y=ndcg_click] {latFactor_click_ndcg.txt};
\addplot[color={rgb:blue,4;green,2;yellow,1}, mark=o, dashed, mark options={solid}]
table[x=para, y=ndcg_buy] {latFactor_click_ndcg.txt};
\legend{\Large Click, \Large Like};

\end{axis}
\end{tikzpicture}

\begin{tikzpicture}
\begin{axis}[
    xlabel={Memory Dimension $M$},
    ylabel={HR},
    xmin=2,xmax=16,
    ymin=0.760,ymax=0.860,
    legend columns=1,
    legend cell align=right,
    grid=both,
    every axis plot/.append style={ultra thick},
    every tick label/.append style={scale=1.3},
    label style={scale=2},
    legend style={
        nodes={scale=1.5, transform shape},
        legend image post style={scale=1.5},
        },
    legend style={at={(0,0)},anchor=south west},
    every axis plot post/.append style={
        every mark/.append style={scale=2}
    }
]

\addplot[color={rgb:red,4;green,2;yellow,1}, mark=otimes, dashed]
table[x=para, y=hr_click] {MemoMat.txt};
\addplot[color={rgb:blue,4;green,2;yellow,1}, mark=o, dashed, mark options={solid}]
table[x=para, y=hr_buy] {MemoMat.txt};
\legend{\Large Click, \Large Like};

\end{axis}
\end{tikzpicture}

\begin{tikzpicture}
\begin{axis}[
    xlabel={Memory Dimension $M$},
    ylabel={NDCG},
    xmin=2,xmax=16,
    ymin=0.480,ymax=0.560,
    legend columns=1,
    legend cell align=right,
    grid=both,
    every axis plot/.append style={ultra thick},
    every tick label/.append style={scale=1.3},
    label style={scale=2},
    legend style={
        nodes={scale=1.5, transform shape},
        legend image post style={scale=1.5},
        },
    legend style={at={(0,0)},anchor=south west},
    every axis plot post/.append style={
        every mark/.append style={scale=2}
    }
]

\addplot[color={rgb:red,4;green,2;yellow,1}, mark=otimes, dashed]
table[x=para, y=ndcg_click] {MemoMat_ndcg.txt};
\addplot[color={rgb:blue,4;green,2;yellow,1}, mark=o, dashed, mark options={solid}]
table[x=para, y=ndcg_buy] {MemoMat_ndcg.txt};
\legend{\Large Click, \Large Like};

\end{axis}
\end{tikzpicture}

\begin{tikzpicture}
\begin{axis}[
    xlabel={Feature Extraction Network Depth $N$},
    ylabel={HR},
    xmin=0,xmax=6,
    ymin=0.780,ymax=0.860,
    legend columns=1,
    legend cell align=right,
    grid=both,
    every axis plot/.append style={ultra thick},
    every tick label/.append style={scale=1.3},
    label style={scale=2},
    legend style={
        nodes={scale=1.5, transform shape},
        legend image post style={scale=1.5},
        },
    legend style={at={(0,0)},anchor=south west},
    every axis plot post/.append style={
        every mark/.append style={scale=2}
    }
]

\addplot[color={rgb:red,4;green,2;yellow,1}, mark=otimes, dashed]
table[x=para, y=hr_click] {depth_click.txt};
\addplot[color={rgb:blue,4;green,2;yellow,1}, mark=o, dashed, mark options={solid}]
table[x=para, y=hr_buy] {depth_click.txt};
\legend{\Large Click, \Large Like};
% \addplot[color={rgb:red,4;green,2;yellow,1}, mark=o, dashed, mark options={solid}]
% table[x=para, y=mae] {embedding_ML10M.txt};

\end{axis}
\end{tikzpicture}

\begin{tikzpicture}
\begin{axis}[
    xlabel={Feature Extraction Network Depth $N$},
    ylabel={NDCG},
    xmin=0,xmax=6,
    ymin=0.460,ymax=0.560,
    legend columns=1,
    legend cell align=right,
    grid=both,
    every axis plot/.append style={ultra thick},
    every tick label/.append style={scale=1.3},
    label style={scale=2},
    legend style={
        nodes={scale=1.5, transform shape},
        legend image post style={scale=1.5},
        },
    legend style={at={(0,0)},anchor=south west},
    every axis plot post/.append style={
        every mark/.append style={scale=2}
    }
]

\addplot[color={rgb:red,4;green,2;yellow,1}, mark=otimes, dashed]
table[x=para, y=ndcg_click] {depth_click_ndcg.txt};
\addplot[color={rgb:blue,4;green,2;yellow,1}, mark=o, dashed, mark options={solid}]
table[x=para, y=ndcg_buy] {depth_click_ndcg.txt};
\legend{\Large Click, \Large Like};

\end{axis}
\end{tikzpicture}
    \end{adjustbox}
    \vspace{-0.2in}
    \caption{Hyper-parameter study in terms of \textit{HR@10} and \textit{NDCG@10}}
    \vspace{-0.15in}
    \label{fig:hyperparam}
\end{figure*}
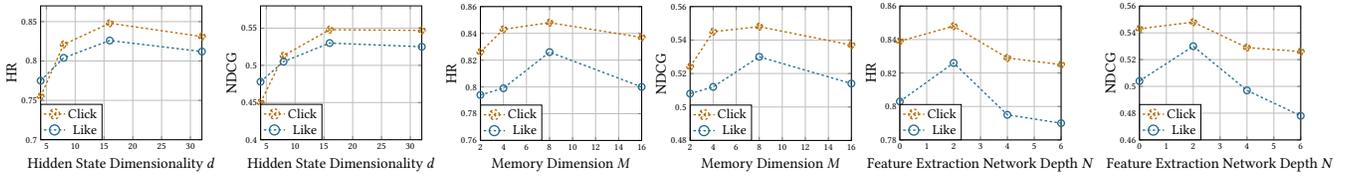

In our experiments, we also investigate the impact of different hyperparameter settings in our developed \emph{\model} framework. Specifically, we evaluate the model recommendation performance by varying the values of several key hyperparameters, including the hidden state dimensionality $d$, the memory dimension $M$ in our memory attention network, and the number of neural network layers $N$ in our deep feature extraction module. The evaluation results on the Yelp data in predicting both click and like behavior are shown in Figure~\ref{fig:hyperparam}. The major findings are summarized as below:
\begin{itemize}[leftmargin=*]
\item \textbf{Hidden State Dimensionality $d$}. We can observe that when we increase $d$ from 4 to 16, the recommendation performance becomes better, but the further increase the $d$ value ($\geq 32$) may not be helpful for the model prediction accuracy. The potential reason for this observation is that the large number of latent units could bring a stronger representation capability.\\\vspace{-0.1in}
\item \textbf{Memory Dimension $M$}. Our memory attention network enables the behavior type-specific semantics learning could be performed from $M$ different dimensions. The parameter study results on memory dimension $M$ indicates that performing the transformation with more latent learning sub-spaces will benefit the recommendation at the early stage, but the continuous increase of $M$ will lead to the overfitting issue.\\\vspace{-0.1in}
\item \textbf{Feature Extraction Network Depth $N$}. We further examine whether designing a deep feature extraction network is beneficial to the recommendation task. As we can see, stacking two hidden layers is beneficial to the performance, which is attributed to the high non-linearities brought by more non-linear layers. However, the overfitting phenomenon can be observe when we perform more transformation-based feature interaction operation with more hidden layers ($\geq 3$).
\end{itemize}

% \begin{table}[h]
% 	\caption{Hyper-parameter study in terms of \textit{HR@10} and \textit{NDCG@10} (\textbf{RQ5})}
% 	\centering
%     \ssmall
%     % \footnotesize
%     % \small
% 	\setlength{\tabcolsep}{0.6mm}
% 	\begin{tabular}{|c|c|c|c|c|c|c|c|c|c|c|c|c|c|}
% 		\hline
% 		\multirow{2}{*}{Data}&\multirow{2}{*}{Metric}&\multicolumn{4}{c|}{Depth of Deep Layers}&\multicolumn{4}{c|}{Latent Factor \#}&\multicolumn{4}{c|}{Memory Matrix \#}\\
% 		\cline{3-14}
% 		&&0&2&4&6&4&8&16&32&2&4&8&16\\
% 		\hline
% 		\hline

%         \multirow{2}{*}{Yelp-Click}&HR
%         &0.839&0.848&0.827&0.825&0.755&0.821&0.848&0.841&0.826&0.843&0.848&0.837\\
%         \cline{2-14}
%         &NDCG&0.543&0.548&0.523&0.526&0.450&0.513&0.548&0.549&0.524&0.545&0.548&0.537\\
% 		\hline
% 		\multirow{2}{*}{Yelp-Buy}&HR
% 		&0.803&0.826&0.795&0.790&0.775&0.804&0.826&0.812&0.794&0.799&0.826&0.800\\
% 		\cline{2-14}
% 		&NDCG&0.504&0.530&0.497&0.492&0.478&0.505&0.530&0.525&0.508&0.512&0.530&0.514\\
% 		\hline
% 	\end{tabular}
% 	\label{tab:ablation}
% 	\vspace{-0.1in}
% \end{table}

\subsection{Case Study on Model Interpretation (RQ5)}
\label{sec:interpretation}
In this subsection, we perform qualitative analyses to show the model interpretation of \emph{\model} in comprehending user behavior relationships and generate more convincing recommendation. To be specific, we visualize the learned quantitative weights learned by our multi-head self-attention mechanism, memory-augmented attention network and multiplex relation aggregation layer. Four typical cases (\ie, samp$_1$,...,samp$_4$) are sampled from the prediction of overall click behavior and purchase behavior on the E-Commerce dataset. From the visualization results, we have the following observations:

(1) \textit{page view} and \textit{purchase} behavior could provide more informative signals in predicting the \emph{click} and \emph{purchase}, respectively. This make sense since the same type of behavior may share closer relationships than other behavior types. (2) In the head-specific self-attention layer, the $4\times 4$ behavior relevance matrix indicates across four types of user behavior. An interesting observation is that: add-to-favorite activity is more like to be correlated with page view and purchase, than add-to-favorite. Similar results can be observed for add-to-cart. It might indicate that add-to-favorite has a high co-occurrence probability than others in the real-world e-commerce platform. (3) The memory attention could learn weights in an adaptive way which corresponds to the importance score generated by our gating mechanism. The reason lies in the utilization of ReLU activation in the attention calculation instead of the mandatory restriction with Softmax function. Overall, all above observations demonstrate the model interpretation power of \emph{\model} in capturing complex behavior relations from different perspective.

\begin{figure}
    \centering
    \includegraphics[width=0.99\columnwidth]{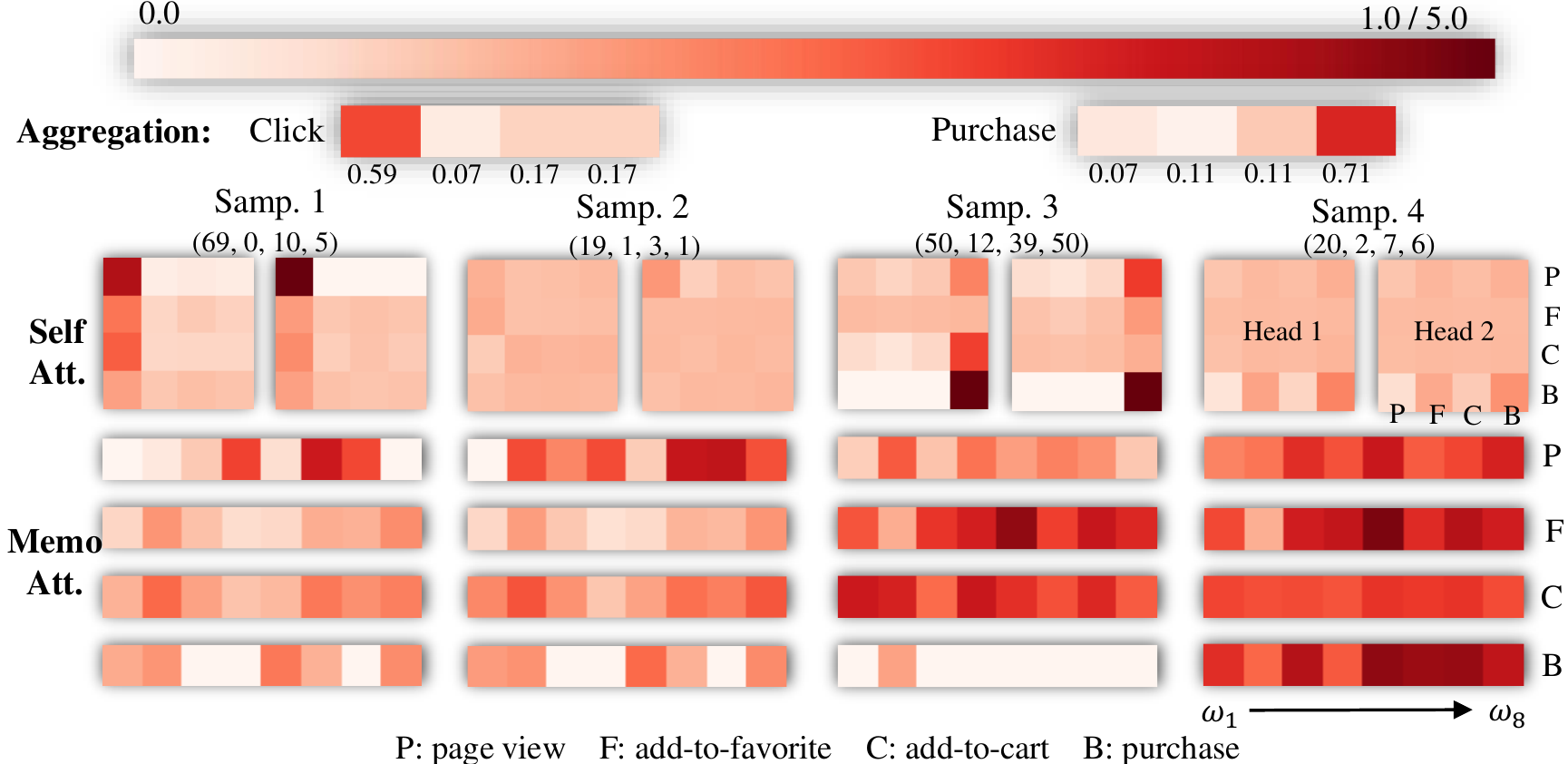}
    % \vspace{-0.1in}
    \caption{Case study of the learned quantitative weights from key modules in \emph{\model}. Pair-wise relations between four types of behavior (\eg, $P$, $F$, $C$ and $B$) are represented with a $4\times 4$ weight matrix in the multi-head self-attention layer. $\omega_1$,...,$\omega_8$ indicate the learned weights across 8 memory dimensions in the memory attention network. The four relevance scores encoded by gating mechanism corresponds to four behavior types. Tuples (\eg, <50,12,39,50>) are numbers of behaviors in the order of (P, F, C, B).}
    \label{fig:case_study}
    % \vspace{-0.1in}
\end{figure}

% 

% To show the interpretation ability, we perform case study on our \model\ by visualizing learned weights from the multi-head attention, memory attention and the weighted aggregation gate. Typical samples drawn from the E-Commerce dataset are shown in Figure~\ref{fig:case_study}, on which we make the following observation: (1) \textit{page view} and \textit{purchase} behavior are the most important in the aggregation layer, for overall prediction and purchase prediction respectively. (2) In the self-attention, other behaviors tend to build strong dependencies with the two aforementioned popular behaviors in their respective prediction tasks. (3) The memory attention learns weights with adaptive scale for different data instances and the magnitude seems to be influenced by the number of categorical behaviors, which we owe to the use of ReLU activation in the attention calculating.

% \vspace{-0.1in}
\subsection{Scalability Study of \emph{\model} (RQ6)}
\label{sec:scalability}
In addition to recommendation accuracy, the model efficiency is also an important factor to investigate. In this subsection, we evaluate the computation time cost of our \emph{\model} as compared to other baselines. In Table~\ref{tab:time}, we report the running time of each epoch during the training phase of each compared approach. We can observe that our \emph{\model} model could achieve comparable performance when competing with most baselines, especially in dealing with the large-scale user-item interaction data.

Although we lose in the cases when comparing with some of the competitive baselines--learning user-item interaction representations with simple relation encoders (\eg, Multilayer Perceptron, vanilla autoencoder), our \emph{\model} still exhibits competitive model scalability due to the comparable time complexity. However, \emph{\model} can show obvious performance superiority over these techniques. In addition, the performance gap (measured by running time) between \emph{\model} and graph neural network recommendation methods (\ie, ST-GCN and NGCF), may stem from the high computational cost of graph convolution operation when performing information aggregation and propagation.

\begin{table}[h]
% \vspace{-0.1in}
    \caption{Computational time cost (seconds) investigation.}
    \centering
    \footnotesize
    % \small
    \vspace{-0.1in}
    \begin{tabular}{lccc}
         \toprule
         Models& Yelp\ \ \  & MovieLens & E-Commerce\\
         \midrule
         BiasMF&34&34&241\\
         DMF&48&59&231\\
         NCF-N&33&35&247\\
         AutoRec&32&37&228\\
         CDAE&33&37&228\\
         CF-NADE&32&37&229\\
         CF-UIcA&51&63&476\\
         ST-GCN&55&65&491\\
         NGCF&61&70&503\\
         NMTR&33&35&268\\
         DIPN&99&134&1062\\
         \hline
         \emph{\model} &40&49&234\\
         \bottomrule
    \end{tabular}
    \vspace{-0.15in}
    \label{tab:time}
\end{table}

\vspace{-0.05in}
\section{Related Work}
\label{sec:relate}

% In this section, we discuss the related work from the following three perspectives which are relevant to this study. \\\vspace{-0.1in}

\noindent \textbf{Deep Collaborative Filtering Techniques}.
Deep learning have been revolutionizing collaborative filtering techniques and achieve promising results in many recommendation scenarios. For example, Multi-layer Perceptron has been integrated into the collaborative filtering architecture to handle non-linear feature interactions~\cite{he2017neuralncf,xue2017deep}. Several work attempts to utilize encoder-decoder network to map explicit user-item interactions into latent representations, using autoencoder~\cite{sedhain2015autorec} and its variants~\cite{strub2016hybrid}. In addition, another research line lie in leveraging graph neural network to incorporate user-item graph signals into the recommendation framework, such as NGCF~\cite{wang2019neural}, STAR-GCN~\cite{zhang2019star} and Multi-GCCF~\cite{sun2019multi}. The major difference between these methods and ours is that \model\ explores the cross-behavior interactive knowledge to assist recommendation. \\\vspace{-0.1in}

\noindent \textbf{Relation-aware Recommender Systems}. Prior work has made significant advances to develop recommender systems with the consideration of various relations between users and items. For example, the social-aware recommender systems aim to boost recommendation performance by exploring user's social relations based on the information dissemination~\cite{fan2019graph,chen2019social}. Furthermore, knowledge graph has become another information source from item side to help recommendation models capture relationships between items~\cite{wang2019knowledge,wang2019kgat}. In addition to the relation of collaborative similarity, there exist work aiming to consider multiple item relationships (\eg, shared director or categories) to learn fine-grained item knowledge~\cite{xin2019relational}. Different from these methods which focus on using the exogenous information from either user or item side, this work explores the multiplicity of pairwise user-item interactions and carefully learns their underlying inter-dependencies. \\\vspace{-0.1in}

\noindent \textbf{Attention Network for Recommendations}. Attention mechanism has been proven to be effective in differentiating various relations for recommendations~\cite{tay2018multi}, such as item transitions~\cite{li2017neural}, user connections~\cite{song2019session} and customer group dynamics~\cite{vinh2019interact}. To address the limitation of recurrent neural architectures in capturing long-range dependencies without the rigid order assumption, self-attention mechanism has been introduced to model correlations from any pair of positions of input data points~\cite{vaswani2017attention}. For example, Sun~\etal~\cite{sun2019bert4rec} proposed a bidirectional self-attention framework for sequential recommendation. Additionally, multi-head self-attentive model is introduced to recommended news to users~\cite{wu2019neural}. Our \model\ framework is motivated by the multi-head self-attentive learning architecture in a sense that a memory augmented transformer is designed to model multiplex behavior relation dynamics from different types of user-item interactions.
\vspace{-0.05in}
\section{Conclusion}
\label{sec:conclusion}

In this work, we propose \model, a novel memory augmented transformer neural architecture which incorporates multiple types of user behavior relationships into a cross-behavior collaborative filtering framework. We argue that these different types of user-item interactions are usually neglected in conventional methods. \model\ demonstrates the state-of-the-art performance on two benchmark datasets and a large-scale user behavior data from a major online retailing platform. In addition, via the qualitative analysis of the attentive weights, we discover that the implicit cross-type behavioral dependencies are encoded within the \model\ framework.

Notwithstanding the interesting problem and promising results, some directions exist for future work. We will next incorporate rich auxiliary data source (\eg, user review text information or item description~\cite{zheng2017joint}) to further enhance the current recommendation framework. Additionally, another time dimension of the problem deserves more investigation. When multi-type user-item interaction data arrives in a timely manner, how to best account for it in the current \model\ framework? One possible direction is adapting~\model\ to a time-sensitive model by analyzing the trade-off between accuracy and complexity.

\section*{Acknowledgments}
We thank the anonymous reviewers for their constructive feedback. This work was supported by National Nature Science Foundation of China (61672241, U1611461), Major Project of National Social Science Foundation of China (18ZDA062), Natural Science Foundation of Guangdong Province (2016A030308013), Science and Technology Program of Guangdong Province (2019A050510010), and Fundamental Research Funds for the Central Universities (x2js-D2192830).

% \clearpage

\bibliographystyle{ACM-Reference-Format}
\bibliography{sigproc}

\end{document}